\documentclass{article}
\usepackage[cmfonts,ninepoint]{eusipco}
\usepackage{xspace}
\usepackage{amsfonts,amssymb,amsbsy,amsmath,euscript}
\usepackage[T1]{fontenc}
\usepackage[francais]{babel}
\usepackage[dvips]{graphicx}

\def\trace{\mathrm{tr}}
\def\inv{^{-1}}
\def\adj{^\dagger }

\def\vect{\mathrm{vect}}

\title{Blind component separation in wavelet space.\\ Application to 
CMB analysis.}
\name{ Y.~Moudden$^1$,  J.-F.~Cardoso$^2$,  J.-L.~Starck$^1$ and  J.~Delabrouille$^3$ }
\address{ $^1$ DAPNIA/SEDI-SAP, CEA/Saclay, F-91191 Gif-sur-Yvette, France\\
$^2$ CNRS/ENST, 46 rue Barrault, F-75634 Paris, France\\
$^3$ CNRS/PCC, Coll\`ege de France, 11 place Marcelin Berthelot, F-75231 Paris, France\\\\
yassir.moudden@cea.fr, cardoso@tsi.enst.fr, jstarck@cea.fr, delabrouille@cdf.in2p3.fr\\}

\begin{document}

\maketitle

\begin{abstract}
It is a recurrent issue in astronomical data analysis that observations are unevenly sampled or  incomplete maps with missing patches or intentionaly masked parts. In addition, many astrophysical emissions are non stationary processes over the sky. Hence spectral estimation using standard Fourier transforms is no longer reliable. Spectral matching ICA (SMICA) is a source separation method based on covariance matching in Fourier space which is successfully used for the separation of diffuse astrophysical emissions in Cosmic Microwave Background observations.  We show here that wavelets, which are standard tools in processing non stationary data, can profitably be used to extend SMICA. Among possible applications, it is shown that gaps in data are dealt with more conveniently and with better results using this extension, wSMICA,  in place of the original SMICA. The performances of these two methods are compared on simulated CMB data sets, demonstrating the advantageous use of wavelets.

 \end{abstract}

\begin{small}
{\it \textbf{ Keywords} : blind source separation, cosmic microwave background, wavelets, data analysis}
\end{small}

\section{Introduction}

The detection of Cosmic Microwave Background (CMB) anisotropies on the sky has been over the past three decades subject of intense activity in the cosmology community. 

The CMB, discovered in 1965 by Penzias and Wilson, is a relic radiation emitted some 13 billion years ago, when the Universe was about 370.000 years old. Small fluctuations of this emission, tracing the seeds of the primordial homogeneities which gave rise to present large scale structures as galaxies and clusters of galaxies, have been observed by a number of experiments such as 
Archeops \cite{2002APh....17..101B}, 
Boomerang \cite{2000Natur.404..955D},
Maxima \cite{2000ApJ...545L...5H} 
and WMAP \cite{2003ApJS..148....1B}
.

The precise measurement of these fluctuations is of utmost importance for Cosmology.  Their statistical properties (spatial power spectrum, Gaussianity) strongly depend upon the cosmological scenarios describing the properties and evolution of our Universe as a whole, and thus permit to constrain these models as well as to measure the cosmological parameters describing the matter content, the geometry, and the evolution of our Universe \cite{jungman96}.  

Accessing this information, however, requires disentangling in the data the contribution of several distinct astrophysical sources, all of which emit radiation in the frequency range used for CMB observations \cite{fb-rg99}. This problem of component separation, in the field of CMB studies, has thus been the object of many dedicated studies in the past.\\

To first order, the total sky emission is modelled as a linear mixture of a few independent processes. 
The observation of the sky with detector $d$ is then a noisy linear mixture of $N_c$ components:
\begin{equation}
  y_{d}(\theta, \phi) 
  = 
  \sum_{j=1}^{N_c} A_{d j} s_{j}(\theta, \phi) + 
  n_{d}(\theta, \phi)
  \label{eq:mixture}
\end{equation}
where $s_{j}$ is the emission template for the $j$th astrophysical process, herein referred to as a \emph{source} or a \emph{component}. The coefficients $A_{dj}$ reflect emission laws while $n_d$ accounts for noise. 
When $N_d$ detectors provide independent observations, this equation can be put in vector-matrix form:
\begin{equation}
X(\theta, \phi) =AS(\theta, \phi)+N(\theta, \phi)
\label{model1}
\end{equation}
where $X$ and $N$ are vectors of length $N_d$, $S$ is a vector of length $N_c$, and A is the $N_d\times N_c$ mixing matrix.
 
Given the observations of such a set of independent detectors, component separation consists in recovering estimates of the maps of 
the sources $s_{j}(\theta, \phi)$. 
Explicit component separation has been investigated first in CMB applications by \cite{1996MNRAS.281.1297T}, 
\cite{fb-rg99}, and \cite{1998MNRAS.300....1H}. In these
applications, recovering component maps is the primary target, and all the parameters of the model (mixing matrix $A_{dj}$, noise levels, statistics of the components, including 
the spatial power spectra) are assumed to be known and used as priors to invert the linear system.

Recent research has addressed the case of an imperfectly known mixing matrix.  It is then necessary, to estimate it (or at 
least some of its entries) directly from the data.  For instance, Tegmark \textit{et
al.} assume power law emission spectra for all components except CMB and SZ, and fit spectral indices to the observations 
\cite{2000ApJ...530..133T}. More recently, blind source
separation or independent component analysis (ICA) methods have been implemented specifically for CMB studies. The work of 
\cite{Bac2000}, further extended by \cite{Maino2002} implements a blind source separation
method exploiting the non--Gaussianity of the sources for their separation, which permits to recover the mixing matrix $A$ and the maps of the sources.
  
Delabrouille et al. \cite{Del2003} propose an approach exploiting the spectral diversity of
components, with the new point of view that spatial power spectra are actually the main unknown 
parameters of interest for CMB observations. The estimation of a set of parameters of the model, among which the spatial power spectra of the components, is made using a set of band-averaged spectral covariance matrices in Fourier space.

While working in the Fourier domain has a number of advantages, it also has a number of drawbacks. When components or noise  are strongly non-stationary, one may wish to avoid the averaging induced by Fourier transforms. In addition, 
when dealing with real-life observations, quite often the coverage is incomplete for a reason or another. Either the 
instrument observes only a fraction of the sky, or some regions of the sky have to be rejected due to localised strong 
astrophysical sources of contamination : compact radiosources or galaxies, strong emitting regions in the galactic plane.

Blind component separation (and in particular estimation of the mixing matrix), as discussed by Cardoso \cite{3easy}, can be achieved in several different ways. The first of these exploits non-Gaussianity of all but possibly one components. However, this is not recommended for mixtures where one component is close to Gaussian and all observations suffer from additive Gaussian noise. The component separation method of Baccigalupi \cite{Bac2000} and Maino \cite{Maino2002} is based on this method. The second, which exploits spectral diversity (or non-stationarity in Fourier domain), has the advantage that detector--dependent beams can be handled easily, since the convolution with a point spread function in direct space becomes a simple product in Fourier space. SMICA is an extension of this approach to noisy observations. Finally, the third method exploits non-stationarity in real space. It is adapted to situations where components are strongly non-stationary in real space.

As an extension of these last two methods, it is natural to investigate the possible benefits of exploiting both non-stationarity and spectral diversity for blind component separation using wavelets. Indeed wavelets are powerful tools in revealing the spectral content of non-stationary data.  In what follows, we first recall in section  \ref{smica} the fundamental principles of Spectral Matching ICA. Then, after a brief reminder of the \emph{\`a trous} wavelet transform,  we discuss in section \ref{wavelets} the extension of SMICA for component separation in wavelet space in order  to deal with non-stationary data. Considering  the problem of incomplete data as a model case of practical significance for the comparison of SMICA and its extension wSMICA,  numerical experiments and results are reported in section  \ref{simulations} . From these, conclusions are drawn in section \ref{conclusion}.

\section{SMICA}\label{smica}

This paragraph recalls the main hypotheses and equations of the 
SMICA algorithm 
which we actually extended to deal with gapped data. For ease of 
presentation, we concentrate on the 1D case since the extension to two dimensional 
data is straightforward. Detailled 
descriptions and discussions of this method can be
found in \cite{Pha2, Car} and results of previous applications to CMB 
analysis can be read in \cite{Del2003, Pat}.  

\subsection{Model and cost function}
Spectral matching ICA is a blind source separation technique that 
overcomes the inseparability of Gaussian sources using standard ICA methods by
relying on their assumed spectral diversity : SMICA allows us to 
recover independent Gaussian colored sources from 
observed noisy mixtures provided their spectra are substantially not proportional 
\cite{Car2}.\\

Considering the linear instantaneous mixing model with additive noise
defined by (\ref{model1}), with the assumption that noise and 
source processes are centered, stationary and independent, and denoting 
$R_{X}(\nu)$, $R_{S}(\nu)$ and $R_{N}(\nu)$ the spectral covariances
of $X$, $S$ and $N$ respectively, it follows from (\ref{model1}) that for any value of the reduced 
frequency $\nu \in [-0.5, 0.5]$,
\begin{equation}
R_{X}(\nu) = A R_{S}(\nu) A ^\dagger + R_{N}(\nu)
\label{model2}
\end{equation}
when we further assume independence between source and noise 
processes. Clearly, independence also implies that  $R_{S}(\nu)$ and $R_{N}(\nu)$ are 
diagonal matrices. \\

Given  a batch of $T$ regularly spaced experimental data samples $X_{t=1\rightarrow T}$ 
and a set $\{ \nu_{q, q=1\rightarrow Q} \}$ of Q different reduced 
frequencies chosen \emph{a priori }, estimates $\widehat{R}_{X}(\nu_q)$ of $R_{X}(\nu_q)$ of the spectral covariance at these frequencies
can be computed easily in a number of ways. The basic idea of spectral matching is to fit the model covariances of equation  (\ref{model2}) 
to these experimental covariances by minimizing, over all or a subset of the model parameters $\theta = \{ R_{S}(\nu_q), R_{N}(\nu_q), A \}$, the functional
\begin{equation}
\phi (\theta) =  \sum _{q=1}^{Q}  \alpha_q \mathcal{D} \Big( \widehat{R}_{X}(\nu_q), A 
R_{S}(\nu_q) A^{\dagger} +  R_{N}(\nu_q) \Big)
\label{Cost1}
\end{equation}
where $\mathcal{D} (.,.) $ is a measure of the divergence between two covariance matrices, and $\alpha_q$ are weights which depend on $q$. This adjustment results in estimates
$\widehat{\theta} = \{ \widehat{R}_{S}(\nu_q), \widehat{R}_{N}(\nu_q), \widehat{A} \}$ of the model
parameters and hence enables us to achieve the desired source separation. It is worth highlighting that resorting to covariances highly reduces data dimension, which is of great interest to astrophysical applications where data sets tend to become
very large.  Moreover, it may be argued in the stationary Gaussian case that this reduction is without significant loss of 
information \cite{Del2003}.\\
 
Although any reasonable set of weights $\alpha_q$ and divergence $\mathcal{D} (.,.) $ can be used in (\ref{Cost1})
to assess spectral mismatch, this will affect the statistical properties of the estimated model parameters 
$\widehat{\theta} = \{ \widehat{R}_{S}(\nu_q), \widehat{R}_{N}(\nu_q), \widehat{A} \}$. Deriving a mismatch criterion from higher statistical 
principles such as maximum likelihood should lead to better such estimates.\\

In the SMICA method, the divergence $\mathcal{D}$ used is given by
\begin{equation}
\mathcal{D}_{KL} (R_1, R_2 ) = \frac{1}{2} \Big( \textrm{Tr}(R_1R_2^{-1}) - \textrm{log}\textrm{det}(R_1R_2^{-1}) - m  \Big)
\end{equation}
which actually derives from the Kullback-Leibler divergence between two centered Gaussian distributions with size $m\times m$ 
covariance matrices $R_1$ and $R_2$. Moreover, assuming constant source $R_{S,q}^f$ and noise $R_{N,q}^f$ power spectra, 
over frequency domains $\{F_q\}_{q\in [1, Q]}$ ,
SMICA uses refined unbiased estimates $\widehat{R}_{X,q}^f$ of the mixture covariance matrices $R_{X,q}$ defined by :
\begin{equation}
\hat{R}_{X,q}^f = \frac{1}{n_q}\sum_{p=0,\frac{p}{T}\in F_q}^{T-1} \widetilde{X}(\frac{p}{T})\widetilde{X}(\frac{p}{T})^\dagger
\label{Esti1}
\end{equation}
where $\tilde{X}$ is the discrete Fourier transform of $X$,
\begin{equation}
\widetilde{X}(\nu)= \frac{1}{\sqrt{T}} \sum_{t=0}^{T-1} X(t)e^{-2\pi j \nu t},
\end{equation}
the $F_q$ are non-overlapping domains in $[-1/2, 1/2]$, symmetric with respect to zero, with their positive parts 
centered on $\nu_q$, and $n_q$ is the number of $\frac{p}{T}$
that fall in $F_q$. It follows from this definition that the entries of $\widehat{R}_{X,q}^f$ are in fact
all real. The statistical grounds and implications of these choices are explored in \cite{Pha2, Car2} where
it is shown that SMICA can be derived asymptotically from the maximum likelihood principle in the
particular case of stationary processes in the Whittle approximation. This latter approximation asserts that the
Fourier coefficients $\widetilde{X}(\frac{p}{T})$ of a stationary process $X(t)$ are asymptotically Gaussian, uncorrelated, centered
with spectral covariance equal to $R_X(\frac{p}{T})$.

As a result, the model covariance (\ref{model2}) is finally rewritten as:
\begin{equation}
 R_{X,q}^f = A R_{S,q}^f A^{\dagger} +  R_{N,q}^f
\end{equation}
and the derived  spectral matching criterion is given by 
\begin{equation}
\phi(\theta) =  \sum _{q=1}^{Q} n_q  \mathcal{D}_{KL} \Big( \widehat{R}_{X,q}^f,
 A R_{S,q}^f A^{\dagger} +  R_{N,q}^f \Big)
\label{Cost2}
\end{equation}
to be minimized with respect to the new set of parameters  $\theta = (A,R_{S,q}^f, R_{N,q}^f )$.\\

The previous definitions are easily extended for the method to be applied to real images.  The above $F_q$ are naturally replaced by 
2D domains in the frequency plane \cite{Car}. These are best chosen, based on available prior information 
relative to source spectra, to enhance spectral diversity \cite{Car2}. Regarding our application to CMB analysis, 
the supposed  spatial stationarity and isotropy of the sources strongly suggests taking rings centered on the null frequency 
which are finally simply described as 1D frequency bands. \\

An especially important limiting case, for simulation purposes, is when the mixing matrix is square
and invertible, and when the mixtures are assumed without noise. Then, as shown in \cite{Pha2}, the likelihood can be directly
related to a joint diagonalization criterion of spectral covariance matrices for which an efficient optimization algorithm 
is actually available.\\

\subsection{Parameter optimization}\label{OPTIM}

Finding the model closest to the data in the sense of SMICA's objective function benefits from the latter's connection 
to the maximum likelihood principle and indeed the EM algorithm is shown to be a fruitful search method in \cite{Del2003} where it
is fully described. Actually, this latter algorithm was slightly modified in order to deal 
with the case of colored noise $N$ in (\ref{model1}). Another useful enhancement was to allow for constraints to be 
set on the model parameters so that prior information such as bounds on some entries of the mixing matrix $A$ could be included.
The details of this constrained EM algorithm are given in appendix \ref{annexe1}.\\ 

Eventually, using the EM algorithm in simulation, it appeared that after a quick start, convergence slowed down dramatically
in a second stage possibly owing to poor signal to noise ratio in some frequency bands. In order to speed convergence back up, 
it was found profitable to alternately use fixed numbers of EM steps and BFGS steps \cite{Del2003, Pat} in a heuristic procedure.\\

An unavoidable issue in optimization is that of initiating the search method and this, obviously, is most critical when the 
function to be optimized is strongly suspected to be multimodal. Such may very well be the case with (\ref{Cost1}). This point though is 
left aside in what follows since our prime interest is in the study of the statistical performances of different estimators
of the model parameters $\theta$. In the simulations discussed further down, the optimal values of the parameters are sought starting
from the true mixing matrix and the spectral covarinces estimated from the initial separate source and noise maps.\\

\subsection{Component map estimation}

As by-products of the SMICA method, estimates $\widehat{R}_{S,q}^f$ and  $\widehat{R}_{N,q}^f$ 
of the different signal and noise covariances are obtained in
the model fitting step and can be used for reconstructing the source maps \emph{via} Wiener filtering
the data maps in Fourier space, in each frequency band $\nu \in F_q$ :
\begin{equation}
 \widehat{S}(\nu) = (\widehat{A}\adj \widehat{R}_{N,q}^{f-1} \widehat{A} + \widehat{R}_{S,q}^{f-1})\inv \widehat{A}\adj \widehat{R}_{N,q}^{f-1} X(\nu)
 \label{wiener}
\end{equation}
In the limiting case where noise is small compared to signal components, $\widehat{R}_{S,q}^{f-1}$ is
negligible and the above filter reduces to 
\begin{equation}
 \widehat{S}(\nu) = (\widehat{A}\adj \widehat{R}_{N,q}^{f-1} \widehat{A})\inv \widehat{A}\adj \widehat{R}_{N,q}^{f-1} X(\nu)
 \label{wiener}
\end{equation}
which is also the generalized least square solution under Gaussian statistics.

Note however that the Wiener filter is only one possibility among others for inverting (\ref{model1}). Its optimality 
is true in the restricted case of Gaussian noise and signal processes. In real case applications,  other inverting schemes should also be experimented \cite{Del2003}.  

\section{Wavelets and SMICA }\label{wavelets}

The SMICA method for spectral matching in Fourier space has proven to be a very powerful tool for CMB spectral estimation in multidetector experiments. It is particularly useful to identify and remove residuals of poorly known correlated systematics and astrophysical foreground emissions contaminating CMB maps. However, SMICA suffers from several practical difficulties when dealing with real data.

Indeed, actual components are known to depart slightly from the ideal linear mixture of equation (\ref{model1}). The mixing matrix (in particular those columns of $A$ which correspond to galactic emissions) is known to depend somewhat on the direction of observation or on spatial frequency. Measuring the dependence $A(\theta,\phi)$ is of interest for future experiments as Planck, and can not be achieved directly with SMICA. Further, the components are known to be both correlated and non stationary.  For instance, galactic dust emissions are strongly peaked towards the galactic plane.  A Fourier (or spherical harmonics) transform inevitably mixes contributions from high galactic sky, nearly free of foreground contamination, and contributions from within the galactic plane.
Noise levels themselves may be quite non stationary, with high SNR regions observed for a long time and low SNR regions poorly observed. 

When there are sharp edges on the maps or  gaps in the data, corresponding to unobserved or masked regions, spectral estimation using the periodogram or the Daniell-like smoothed periodogram 
as in (\ref{Esti1}) is also not the most satisfactory procedure. Although apodizing windows may help cope with edge effects in
Fourier analysis, they are not very straightforward to use in the case of arbitrarily shaped 2D maps with arbitrarily shaped 2D gaps, such as provided by the Archeops experiment \cite{2002APh....17..101B}. Clearly, the spectral analysis of gapped data requires tools different from those 
used to process full data sets, if only because the hypothesized stationarity of the data 
is greatly disturbed by the missing samples.\\

Common such methods often amount to first trying to fill the gaps with estimates of the missing samples 
and then using standard spectral estimators. However, the data interpolation stage is critical 
and cannot be completed without prior assumptions on the data \cite{Sto}. We prefered to rely on 
methods intrinsically dedicated to the analysis of non-stationnary data such as the wavelet transform, widely used to reveal 
variations in the spectral content of time series or images, as they permit to single out regions in direct space while retaining localization in the frequency domain.  We see next how to reformulate (\ref{Cost1}) so to take advantage
of wavelet transforms when dealing with non-stationary data. A particular case in which wavelets are shown to be an especially powerful tool
is that of incomplete data. Note that in what follows, the locations of the missing samples are always known. \\

\subsection{Wavelet transform: the \emph{\`a trous} algorithm}

We give here the necessary background on the \emph{\`a trous} algorithm which, among the several possible wavelet transform implementations, is the one we retained in our simulations. 
With the compact supported cubic $B_3$ spline as scaling function $\phi(k)$, or its 2D quasi-isotropic extension $\phi(k)\phi(l)$, the \emph{\`a trous} algorithm has been shown to be well 
suited to the analysis of atrophysical data where translation invariance is desirable and the accent is seldom set on data 
compression \cite{Sta}. For this choice of scaling function, the scaling equation (\ref{ScalingEquation}) is satisfied and therefore fast 
implementations of the decomposition and reconstruction steps of the \emph{\`a trous} tranform are available \cite{Sta}.\\

Consider for instance a sampled 1D signal $c_0(k)$ of length $T$. The \emph{\`a trous} algorithm recursively produces smoother 
approximations $c_i$ to $c_0$ on a dyadic resolution scale using a low-pass filter $h$ according to :
\begin{equation}
c_i(k) = \sum_{u}h(u)c_{i-1}(k + 2^{i-1} u) = \sum_{u} \frac{1}{2^i}\phi(\frac{k-u}{2^i})  c_0(u)
\end{equation}
where $h = \{ 1/16, 1/4, 3/8, 1/4, 1/16 \}$ is actually the set of coefficients in the scaling equation for the cubic spline :
\begin{equation}
\phi(k) = \sum_u h(u) \phi(2k - u)
\label{ScalingEquation}
\end{equation}
We note that each $c_i$ is the same size as the original data $c_0$ and that the lowest resolution $J_{max}$ is obviously limited by data size $T$. 
Then, taking the difference between two consecutive 
approximations gives the details at that scale or the wavelet coefficients 
\begin{equation}
w_i(k) = c_{i-1}(k) - c_{i}(k) = \sum_{u} \frac{1}{2^{i-1}}\psi(\frac{k-u}{2^{i-1}})  c_0(u)
\end{equation}
where the wavelet function $\psi(k)$ is defined by:
\begin{equation}
\psi(k) = \phi(k) - \frac{1}{2} \phi(\frac{k}{2})
\end{equation}
The $w_i$'s and $c_i$'s given using the \emph{\`a trous} algorithm actually are obtained by passing the
original signal $c_0$ through a set of finite impulse response (FIR) filters $\psi_1, \psi_2, \ldots, \psi_J, \phi_J$. An essential property of 
these filters is that an inverse transform exists. In fact, reconstruction results simply from adding all the wavelet scales 
together with the last smooth approximation:
\begin{equation}
\forall k, c_0(k) = c_J(k) + w_J(k) +w_{J-1}(k) + \ldots + w_2(k) + w_1(k)
\label{IWT}
\end{equation}

The above {\em \`a trous} algorithm is easily extendable to two-dimensional images:
\begin{eqnarray}
c_{i}(k,l) \!\!\!&  = &\!\!\! \sum_u \sum_v h(u,v) c_{i-1}(k+2^{i-1}u,l+2^{i-1}v) \\
w_{i}(k,l) \!\!\!&  = &\!\!\!  c_{i-1}(k,l) - c_{i}(k,l)
\end{eqnarray}
and the reconstruction is still a simple co-addition of the wavelet scales and the smooth array:
\begin{eqnarray}
c_{0}(k,l) = c_{J}(k,l) + \sum_{i=1}^{J} w_{i}(k,l)
\label{eqn_rec}
\end{eqnarray}

The use of the $B_3$ spline leads to a convolution with the $5 \times 5$ mask $h$:
$$
\frac{1}{256}\left(    \begin{array}{ccccc}
1 & 4 & 6 & 4 & 1 \\
4  & 16 & 24  & 16 & 4  \\
6 & 24 & 36  & 24 & 6 \\
4  & 16 & 24  & 16 & 4  \\
1 & 4 & 6 & 4 & 1
\end{array} \right)
$$
but it is faster to compute the convolution in a separable way
(first on rows, and then on the resulting columns).

\subsection{Spectral matching in wavelet space : wSMICA}

Consider the set of ideal band pass filters $\mathcal F_q$ associated with non-overlapping frequency domains 
$F_q$ as used by the Fourier space implementation of SMICA. Let $Y_q$  denote the stationary Gaussian random processes obtained 
by passing  the observations $X$  of size $m$ through filter $\mathcal F_q$. Let $\widetilde{Y}_q$  be their Fourier coefficients. 
Because of the unitary property of the Fourier transform, considering a batch of $T$  samples $X_{t=1,T}$,
the following equality between joint probabilities holds :
\begin{equation}
P(Y_{1;t=1,T},...,Y_{Q;t=1,T} ) = P(\widetilde{Y}_{1;k=1,T},...,\widetilde{Y}_{Q;k=1,T} )
\end{equation}
Assuming uncorrelated Fourier coefficients as in the above mentioned maximum likelihood  derivation of SMICA based  
on the the Whittle approximation,
 and because of the non-overlapping filters, it follows that the $Y_{q;t}$ for different $q$'s are also decorrelated so that:
\begin{equation}
-\mathrm{log} P(Y_{1;t=1,T},...,Y_{Q;t=1,T} ) = -\sum _{q=1}^Q \mathrm{log} P(\widetilde{Y}_{q;k=1,T})
\end{equation}
and that $\forall q$:
\begin{equation}\label{weights}
\begin{split}
 -\mathrm{log} P(Y_{q;k=1,T}) & = -\mathrm{log} P(\widetilde{Y}_{q;k=1,T}) \\
& = n_q  \mathcal{D}_{KL} \Big( \widehat{R}_{X,q}^f, A R_{S,q}^f A^{\dagger} +  R_{N,q}^f \Big)
\end{split}
\end{equation}

Now define mixture, source and noise covariances  $R_{X,q}^t$, $R_{S,q}^t$ and $R_{N,q}^t$  in the time domain
at the output of the above filters. The former matrices can be estimated from the available data using:
\begin{equation}
\widehat{R}_{X,q}^t = \frac{1}{T}\sum_{t=0}^{T-1} Y_{q;t}Y_{q;t}^\dagger
\label{cov_temps}
\end{equation}
and nothing opposes attempting component separation by spectral matching in the time domain using these latter covariances by minimizing
\begin{equation}
\phi (\theta) =  \sum _{q=1}^{Q}  \alpha_q \mathcal{D} \Big( \widehat{R}_{X,q}^t, A 
R_{S,q}^t A^{\dagger} +  R_{N,q}^t\Big)
\label{Cost_time}
\end{equation}
with respect to $\theta = (A,R_{S,q}^t, R_{N,q}^t )$, provided the estimated covariances are full rank matrices. However, deriving adequate weights $\alpha_q$ in order to get 
a good approximation of the likelihood is not straightforward because of the correlations between the $Y_{q;t}$'s at
different $t$'s. In fact, owing to these correlations, the convergence of $\widehat{R}_{X,q}^t$ to $R_{X,q}^t$ can be very slow. The
helpful point equation (\ref{weights}) actually makes is that taking  $\alpha_q = n_q$ will correctly reflect our confidence in 
 the estimated covariances $\widehat{R}_{X,q}^t$.\\

\begin{figure}[!h]
\begin{center}
\includegraphics[width=7cm]{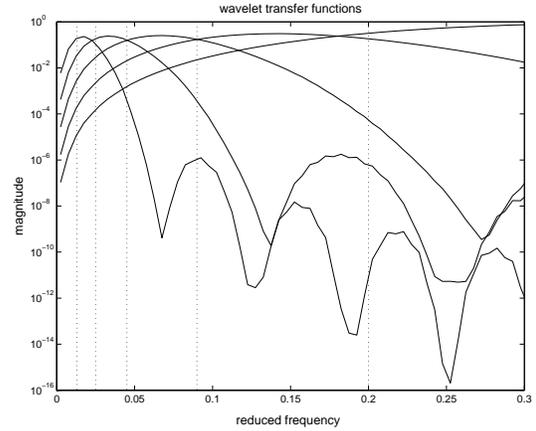}
\caption{Magnitudes of the cubic spline wavelet filters $\psi_1, \psi_2, \ldots, \psi_5$ used in the 
simulations described  further down. The vertical dotted lines for $\nu = \{ 0.013, 0.025, 0.045, 0.09, 0.2, 0.5\}$ 
delimit the five frequency bands used with SMICA in these simulations. }\label{Filtres2}
\end{center}
\end{figure}

The next step is obviously to use another set of filters in place of the ideal band pass filters used by SMICA. In fact, in dealing
with non stationary data or, as a special case, with gapped data, it is especially attractive to consider finite impulse response
filters. Indeed, provided the response of such a filter is short enough compared to data size $T$ and gap widths, not all
the samples in the filtered signal will be affected by the gaps. Therefore, using these latter samples exclusively, one may expect 
better estimation of the statistical properties of the original data \emph{i.e.} without the gaps. We choose
in what follows to use filters $\psi_1, \psi_2, \ldots, \psi_J, \phi_J$ (see figure \ref{Filtres2}) and the wavelet \emph{\`a trous}
algorithm described previously. An immediate consequence of this choice
is that the decorrelation between the different filter outputs
no longer holds, due to their overlapping responses in Fourier space. However, we do benefit from the fast filtering algorithms and,
which is quite significant, from  the possibility of reconstructing estimated source templates.\\

Let us consider again a batch of $T$ regularly spaced data samples $X_{t=1\rightarrow T}$. Possible gaps in the data
are simply described with a mask $\mu$ \emph{i.e.} a vector of zeroes and ones the same length as $X$ with ones corresponding 
to samples outside the gaps. Denoting $W_1, W_2, \ldots, W_J$ and $C_J$ the wavelet scales and smooth 
approximation of $X$, obtained with the \emph{\`a trous} transform and $\mu_1, \ldots, \mu_{J+1}$ the masks for the different scales
determined from the original mask $\mu(t)$ knowing the different filter lengths, wavelet covariances are estimated as follows:
\begin{equation}
\begin{split}
\widehat{R}_{X,1\leq i\leq J}^w = &  \frac{1}{l_i }  \sum_{t = 1}^{T}\mu_i(t) W_i(t) W_i(t) ^\dagger \\
\widehat{R}_{X,J+1}^w = &  \frac{1}{l_{J+1}} \sum_{t = 1}^{T}\mu_{J+1}(t) C_J(t) C_J(t) ^\dagger 
\end{split}
\end{equation}
where $l_i$ is the number of non zero samples in $\mu_i$. With source and noise covariances $R_{S,i}^w, R_{N,i}^w$ defined 
in a similar way, the covariance model in wavelet space becomes 
\begin{equation}
R_{X,i}^w = A R_{S,i}^w A^{\dagger} +  R_{N,i}^w
\end{equation}
and minimizing 
\begin{equation}
\phi (\theta) =  \sum _{q=1}^{Q}  \alpha_q \mathcal{D} \Big( \widehat{R}_{X,q}^w, A 
R_{S,q}^w A^{\dagger} +  R_{N,q}^w\Big)
\label{Cost_wavelet}
\end{equation}
with respect to the model parameters $\theta_w = (A,R_{S,i}^w, R_{N,i}^w )$ achieves the desired component separation.\\
 
However, in order for $\phi (\theta)$ to be a good approximation 
to the likelihood, the weights $\alpha_q$ again have to be determined with care. These weights should account
for the correlations between wavelet coefficients from different or the same scales, especially in the  lower frequencies.  
Actually, exagerating the so-called decorrelating property of the wavelet transform, we assume 
coefficients from different scales are uncorrelated. Nevertheless, coefficients from one same scale are strongly correlated,
especially with the adopted \emph{\`a trous} redundant transform. Then, in the case of complete data sets \emph{i.e.} without gaps,
 and because the 1D wavelet filter length in the time domain
doubles from scale to scale, the transposition of equation (\ref{weights}) leads to taking:
\begin{equation}
\{ \alpha_1, \alpha_2, ..., \alpha_J, \alpha_{J+1} \} = \{ \frac{1}{2}, \frac{1}{4}, ..., \frac{1}{2^J}, \frac{1}{2^J} \}
\end{equation}
In the 2D case, this becomes:
\begin{equation}
\{ \alpha_1, \alpha_2, ..., \alpha_J, \alpha_{J+1} \} = \{ \frac{3}{4}, \frac{3}{16}, ..., \frac{3}{4^J}, \frac{1}{4^J} \}
\end{equation}
However, when there are gaps in the data, the Fourier modes can be strongly correlated and the Whittle approximation
is no longer appropriate. In order to derive an approximate likelihood function, consider the orthogonal discrete
wavelet transform. In the 1D case, this is a non-redundant transform in which the number of coefficients is halved from scale to scale.
It is common and quite convenient to assume these coefficients are uncorrelated. Denoting $l_i^{DWT}$ the number of DWT 
coefficients unaffected by the gaps in scale $i$, these have the same statistical significance or information content
as the $l_i \approx 2^i\times l_i^{DWT}$ coefficients in scale $i$ determined with the \emph{\`a trous} wavelet transform.
Finally, a good approximation to the likelihood is obtained taking 
\begin{equation}
\{ \alpha_1, \alpha_2, ..., \alpha_J, \alpha_{J+1} \} = \{ \frac{l_1}{2}, \frac{l_2}{4}, ..., \frac{l_J}{2^J}, \frac{l_{J+1}}{2^J} \}
\end{equation}
or, in the 2D case,:
\begin{equation}
\{ \alpha_1, \alpha_2, ..., \alpha_J, \alpha_{J+1} \} = \{ \frac{3l_1}{4}, \frac{3l_2}{16}, ..., \frac{3l_J}{4^J}, \frac{l_{J+1}}{4^J} \}
\end{equation}
in equation (\ref{Cost_wavelet}). We will refer to this combination of principles from SMICA and wavelet transforms as wSMICA.\\

A point to be stressed here is that the number of bands in the case of wSMICA is very much limited by
the original data size, which is not as strongly the case with SMICA. But this limitation is mostly a requirement for
reconstruction using (\ref{wiener}) and (\ref{IWT}) to make sense. If the mixing matrix $A$ is a parameter of greater interest and if there is no real 
need to estimate source maps $S$, then there is no objection in principle to using more redundant transforms such 
as the continuous wavelet transform, or in fact any set of linear filters (of finite impulse response 
to cope easily with edges and gaps). This in turn raises the question of optimally 
choosing this set of filters as in \cite{Sto}.

\section{NUMERICAL EXPERIMENTS}\label{simulations}

\subsection{Simulated data}

\begin{figure}[!h]
\begin{center}
\includegraphics[width=5cm]{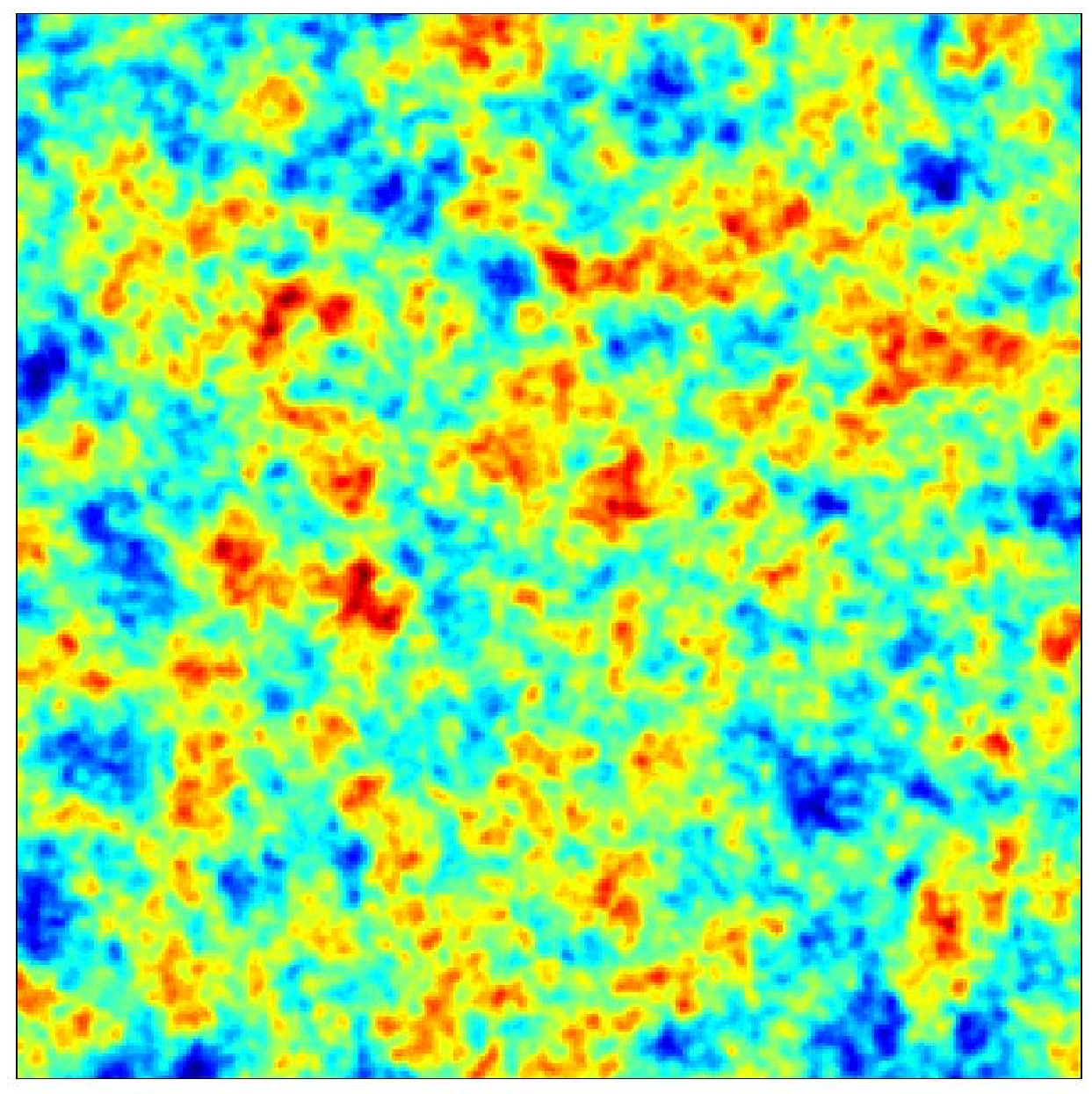}
\\
\includegraphics[width=5cm]{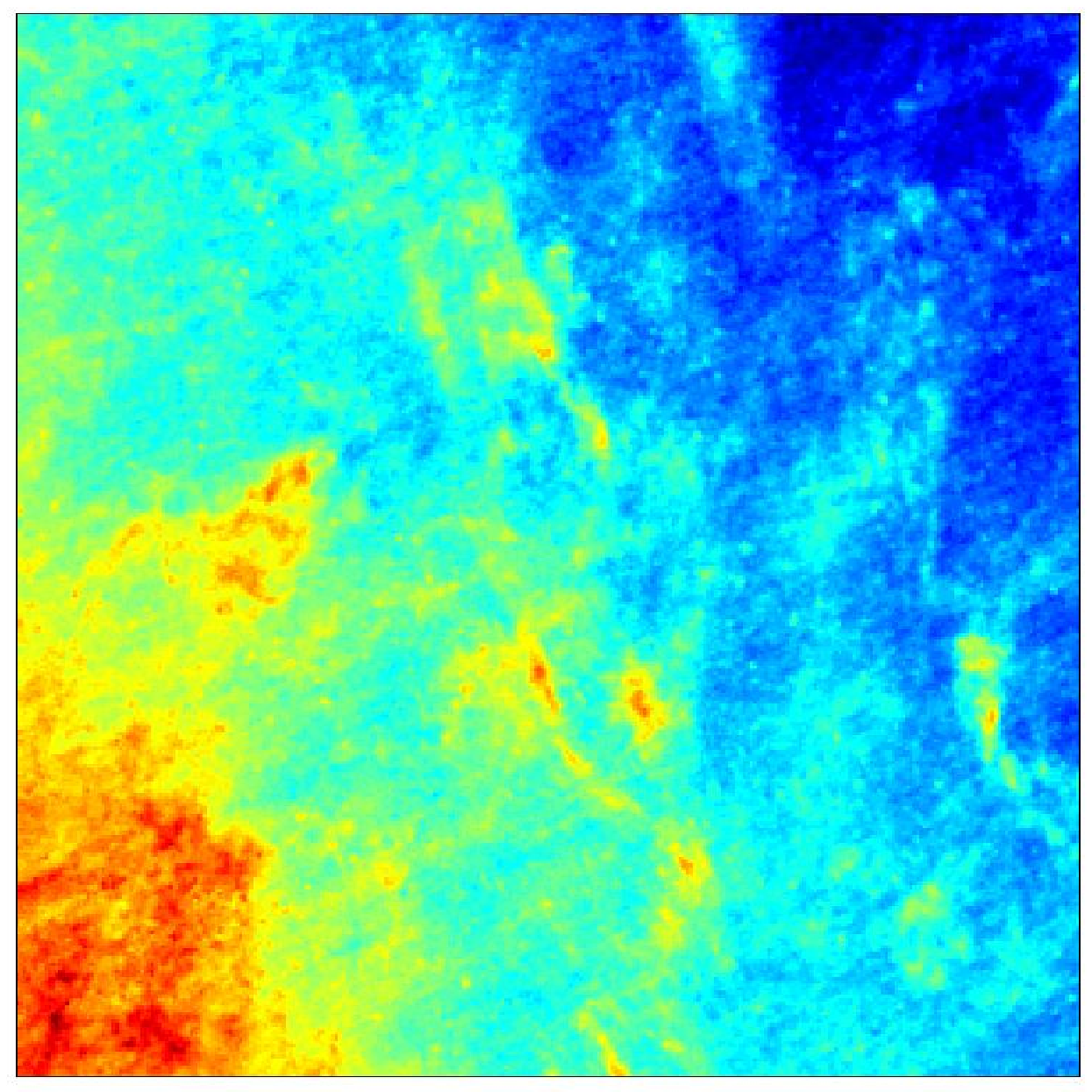}\\
\includegraphics[width=5cm]{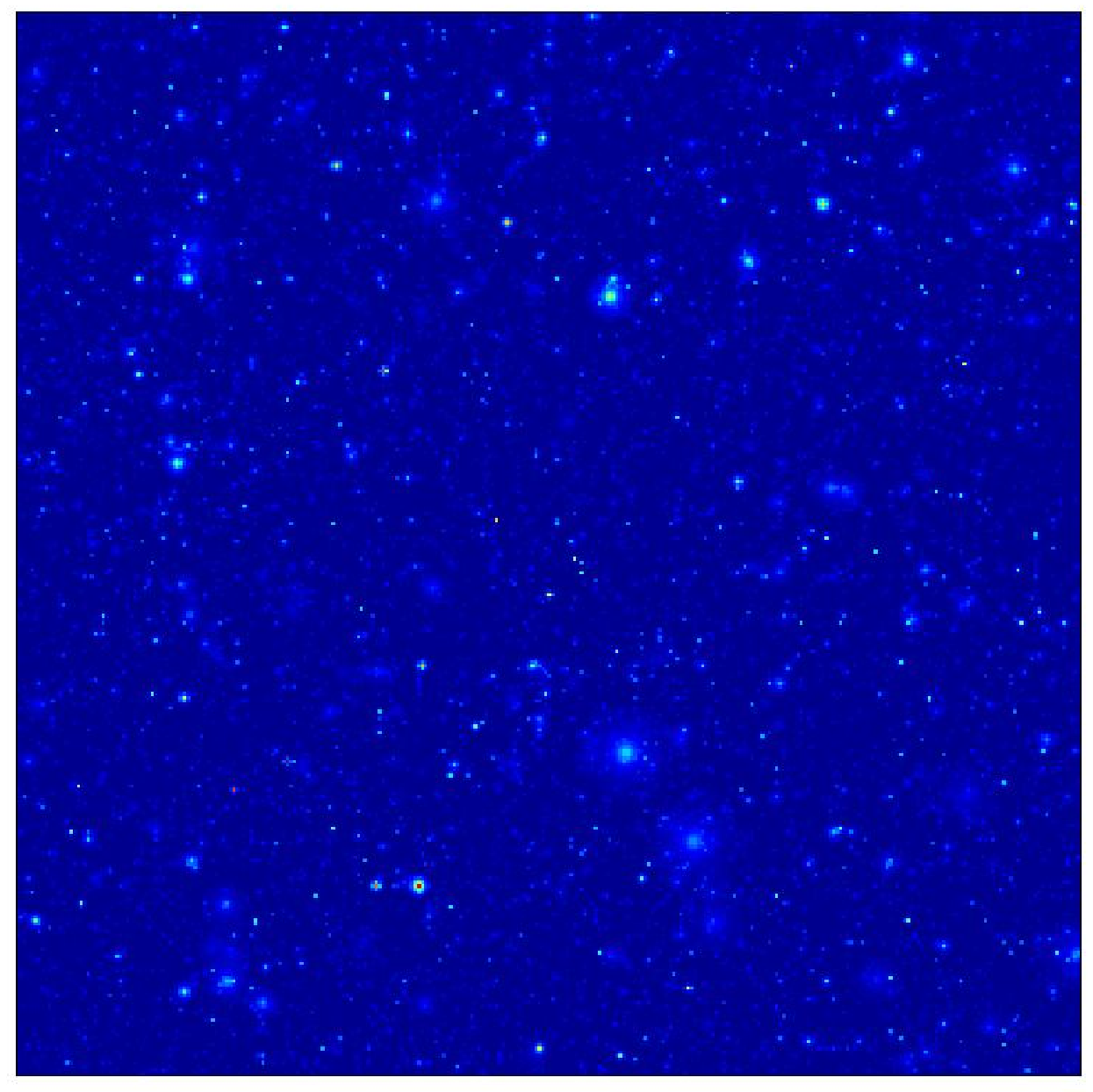}
\\
\caption{Simulated component templates for CMB \emph{(top)}, DUST
\emph{(middle)}, SZ \emph{(bottom)}.} \label{Templates}
\end{center}
\end{figure}

The methods described above were applied to synthetic observations consisting of $m = 6$ mixtures of $n = 3$ 
components namely CMB, galactic dust and SZ emissions for which 
typical templates, shown on figure \ref{Templates}, were obtained 
as described in \cite{Del2003}. \\

The templates, and thus the mixtures in each simulated data set, 
consist of $300 \times 300$ pixel maps 
corresponding to a $12.5^\circ \times 12.5 ^\circ$ field located at 
high galactic latitude. The six mixtures in each set 
mimic observations that will eventually be acquired in the 
six frequency channels of the Planck-HFI on part-sky, local maps. The 
entries of the mixing matrix $A$ used in these simulations 
actually are estimated values of the electromagnetic emission 
laws of the original components at $100$, $143$, $217$, $353$, $545$ and 
$857~\textrm{GHz}$. These values are grouped in 
table \ref{MatrixA}.\\

\begin{table}[!h]
\begin{center}
\tiny{
\begin{tabular}{@{} cccc|c @{}}
CMB & DUST & SZ &  & channel \\
& & & &\\
\hline
& & & &\\
$7.452\times10^{-1}$  &  $ 3.654\times10^{-2}$ &  $-8.733\times10^{-1}$ && 100~GHz\\
$5.799\times10^{-1}$  & $ 7.021\times10^{-2}$ & $-4.689\times10^{-1}$ && 143~GHz\\
$3.206\times10^{-1}$ &  $ 1.449\times10^{-1}$ & $-2.093\times10^{-3}$ && 217~GHz\\
$7.435\times10^{-2}$ &   $3.106\times10^{-1}$&  $ 1.294\times10^{-1}$ & &353~GHz\\
$6.009\times10^{-3}$ &  $5.398\times10^{-1}$ &   $2.613\times10^{-2} $& &545~GHz\\
$6.115\times10^{-5}$  &  $7.648\times10^{-1}$  & $ 5.268\times10^{-4} $& & 857~GHz\\
\end{tabular}}
\caption{Entries of $A$, the mixing matrix used in our simulations.}\label{MatrixA}
\end{center}
\end{table}

\begin{figure}[!h]
\begin{center}
\includegraphics[width=4.2cm]{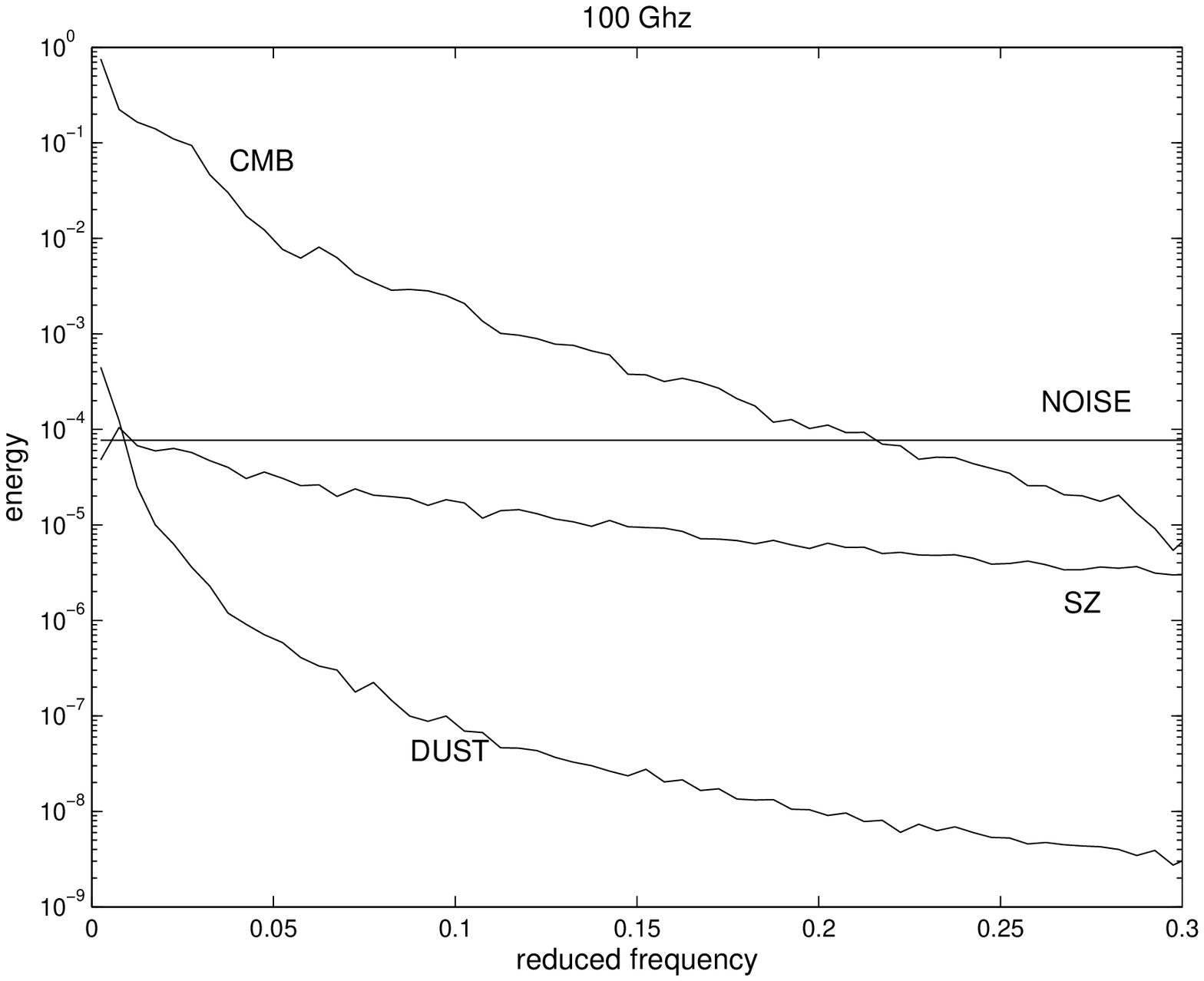}
\includegraphics[width=4.2cm]{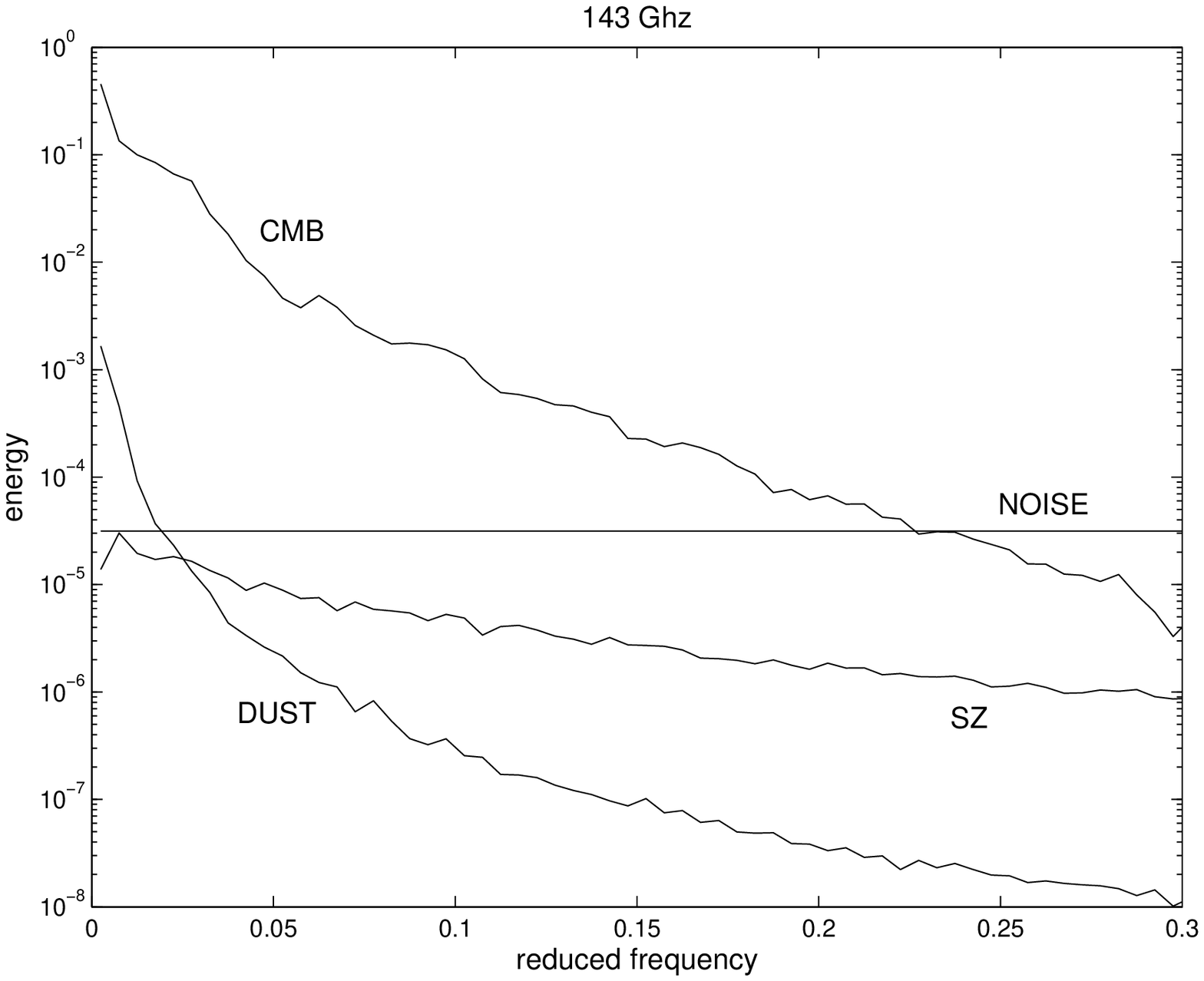}
\includegraphics[width=4.2cm]{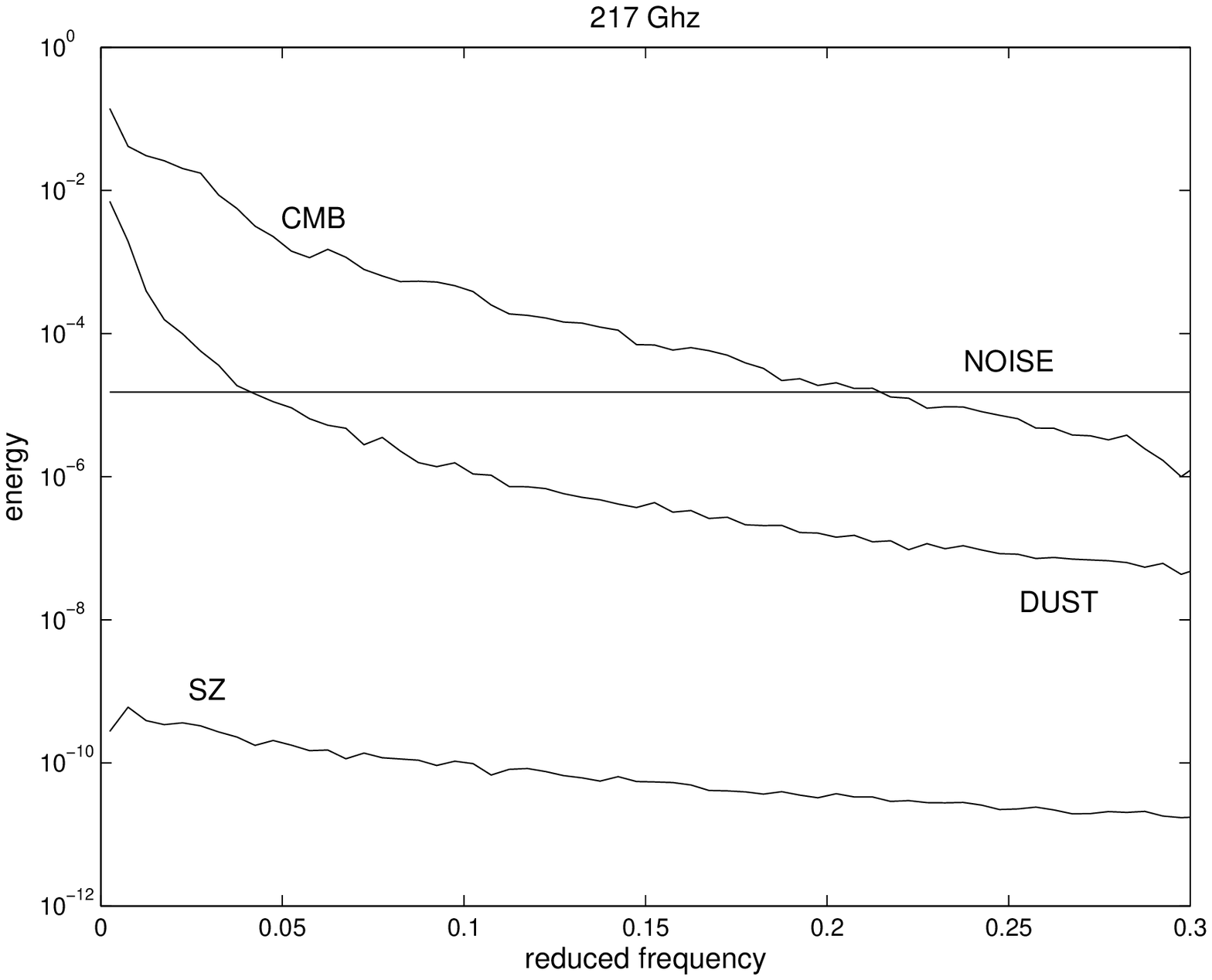}
\includegraphics[width=4.2cm]{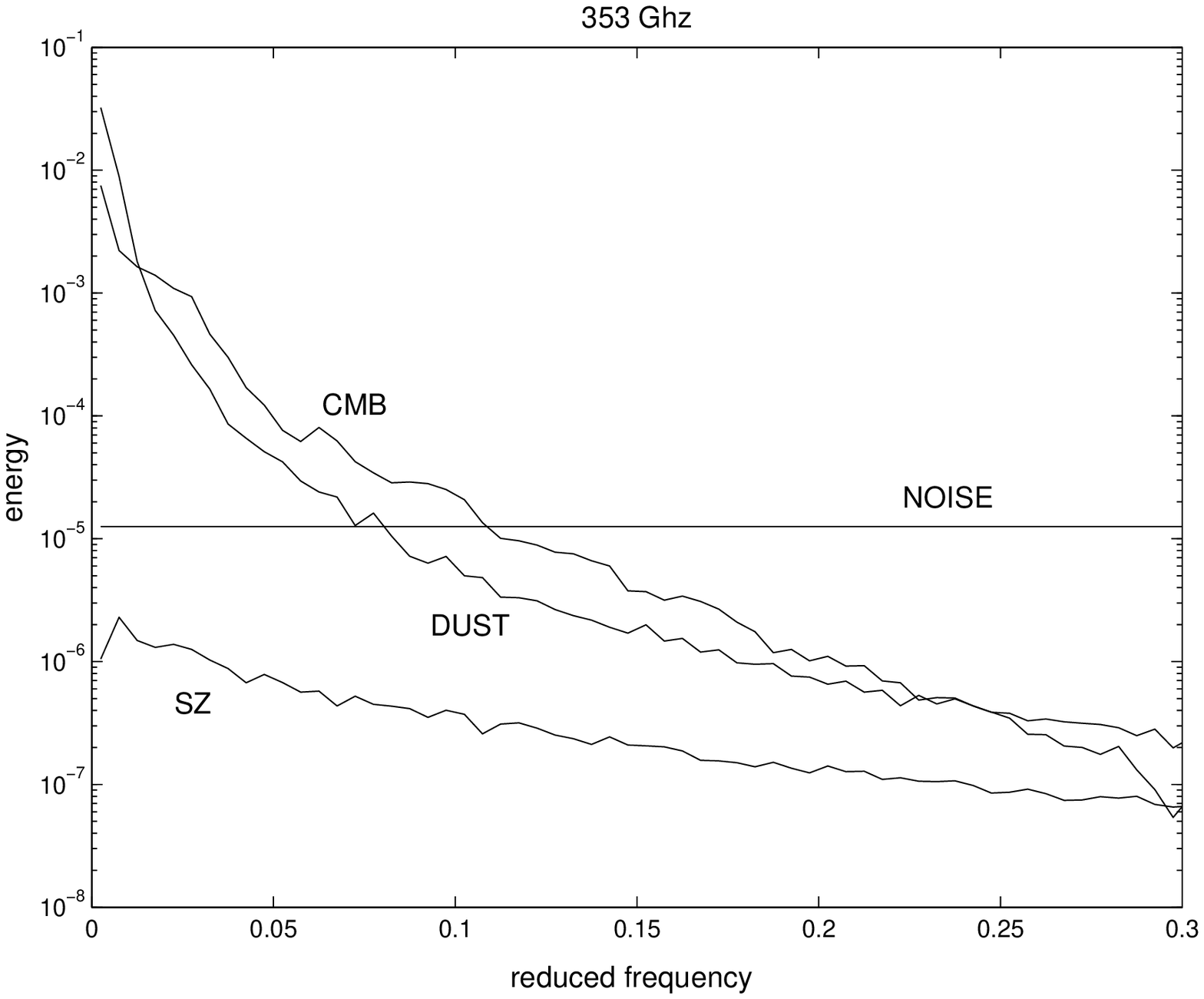}
\includegraphics[width=4.2cm]{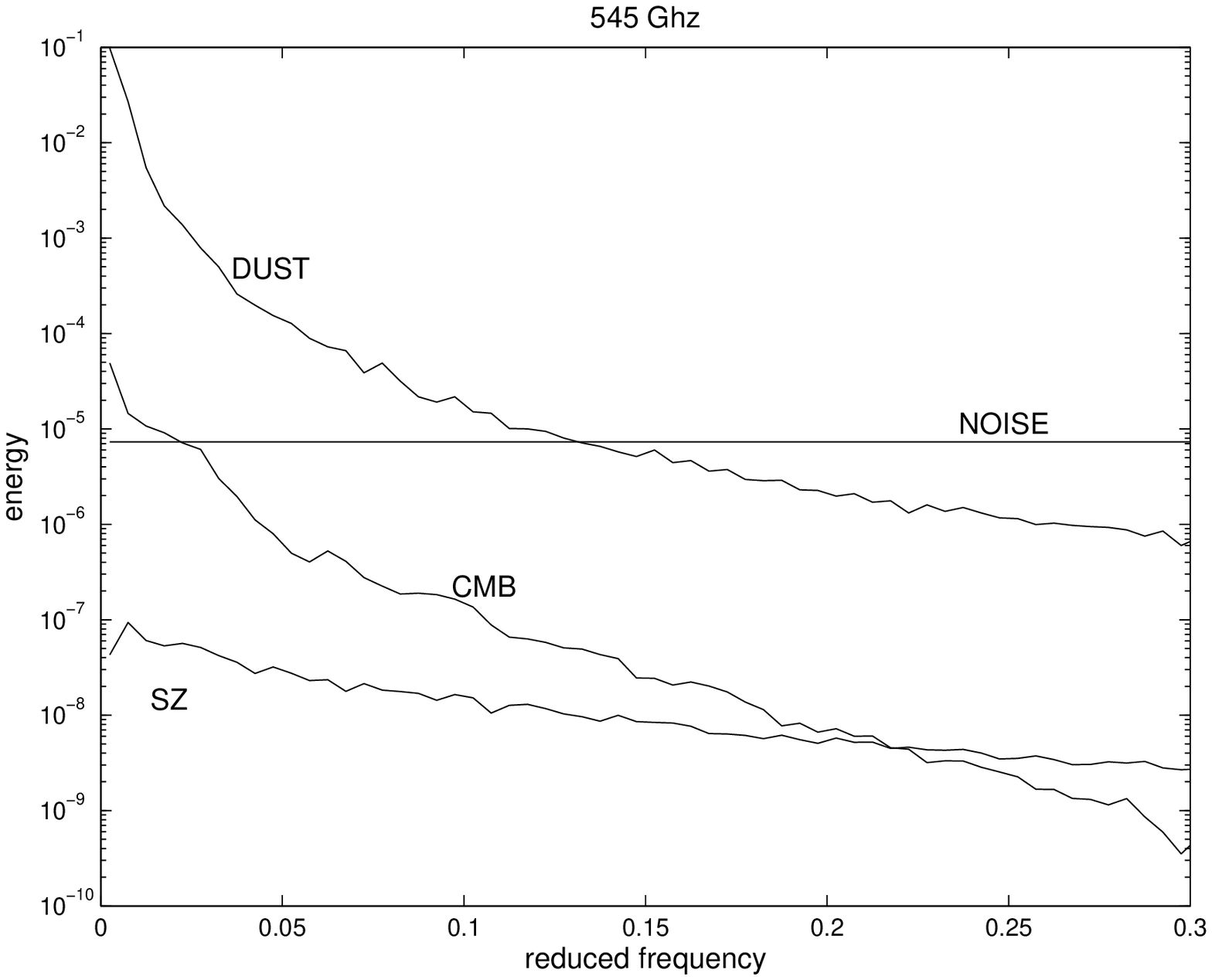}
\includegraphics[width=4.2cm]{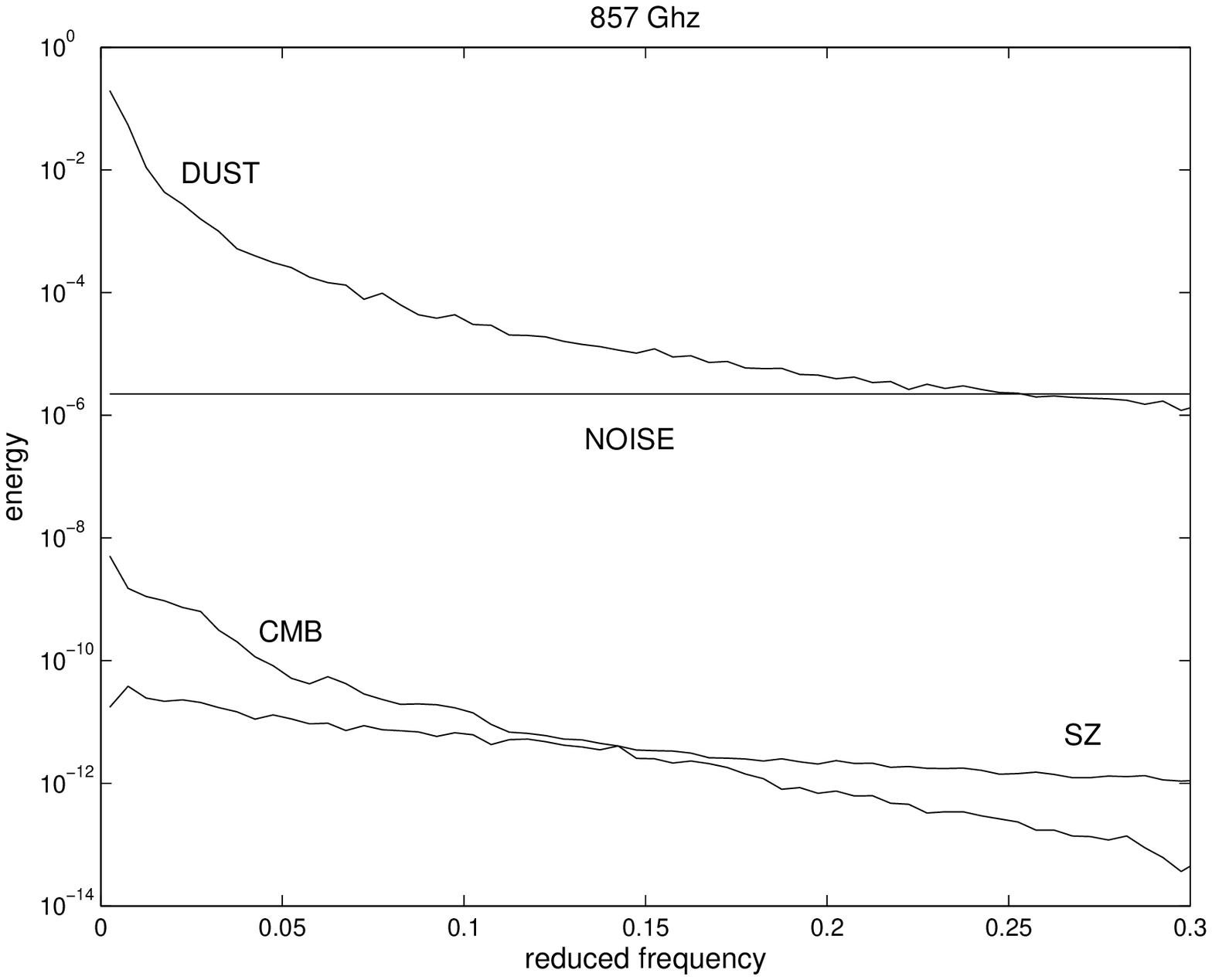}
\caption{Energy contributed by each source and noise  to each of the six mixtures
\emph{ (mixture 1 : top left, mixture 6 : bottom right)}
as a function of frequency, for the nominal noise variance on the Planck HFI channels. Note how SZ is expected to always be below 
nominal noise, that CMB and dust strongly dominate in different channels and that CMB and dust spectra, without being proportional, display the same general behaviour dominated by low modes.}\label{mixtures1}
\end{center}
\end{figure}

White Gaussian noise was added to the mixtures according to equation (\ref{model1}) in order to simulate instrumental noise.
While the relative noise standard deviations between channels were set according to the nominal values of the Planck HFI, we experimented
five \emph{global} noise levels at $-20$, $-6$, $-3$, $0$ and $+3$~dB from nominal values. Table \ref{Energie1} gives the typical
energy fractions that are contributed by each of the $n=3$ original sources and noise, to the total energy 
of each of the  $m=6$ mixtures, considering Planck nominal noise variance. In fact, because SMICA and wSMICA actually work 
on spectral bands, a much better indication of signal to noise ratio in these simulations is given by figure \ref{mixtures1} where it is 
shown how noise and source energy contributions distribute with respect to frequency in the six mixtures. \\

\begin{table}[!h]
\begin{center}
\tiny{
\begin{tabular}{@{} cccc|c @{}}

CMB & DUST & SZ & noise & channel\\
& & & &\\
\hline
& & & &\\
$9.91\times10^{-1}$  &  $ 1.18\times10^{-4}$ &  $7.92\times10^{-3}$ & $2.53\times10^{-6}$& 100~GHz\\
$9.97\times10^{-1}$  & $ 7.25\times10^{-4}$ & $3.79\times10^{-3}$ & $5.17\times10^{-7}$& 143~GHz\\
$9.98\times10^{-1}$ &  $ 1.01\times10^{-2}$ & $2.48\times10^{-7}$ & $1.34\times10^{-7}$& 217~GHz\\
$5.55\times10^{-1}$ &   $4.8\times10^{-1}$&  $ 9.78\times10^{-3}$ &  $7.47\times10^{-8}$&353~GHz\\
$2.5\times10^{-3}$ &  $1.0$ &   $2.75\times10^{-4} $& $3.78\times10^{-9}$ &545~GHz\\
$1.29\times10^{-7}$  &  $1.0$  & $ 5.56\times10^{-8} $& $1.24\times10^{-10}$ & 857~GHz\\

\end{tabular}}
\caption{Energy fraction contributed by each source to the total energy of each 
mixture, for the nominal noise variance on the Planck HFI channels. }\label{Energie1}
\end{center}
\end{table}

\begin{figure}[!h]
\begin{center}
\includegraphics[width=3cm]{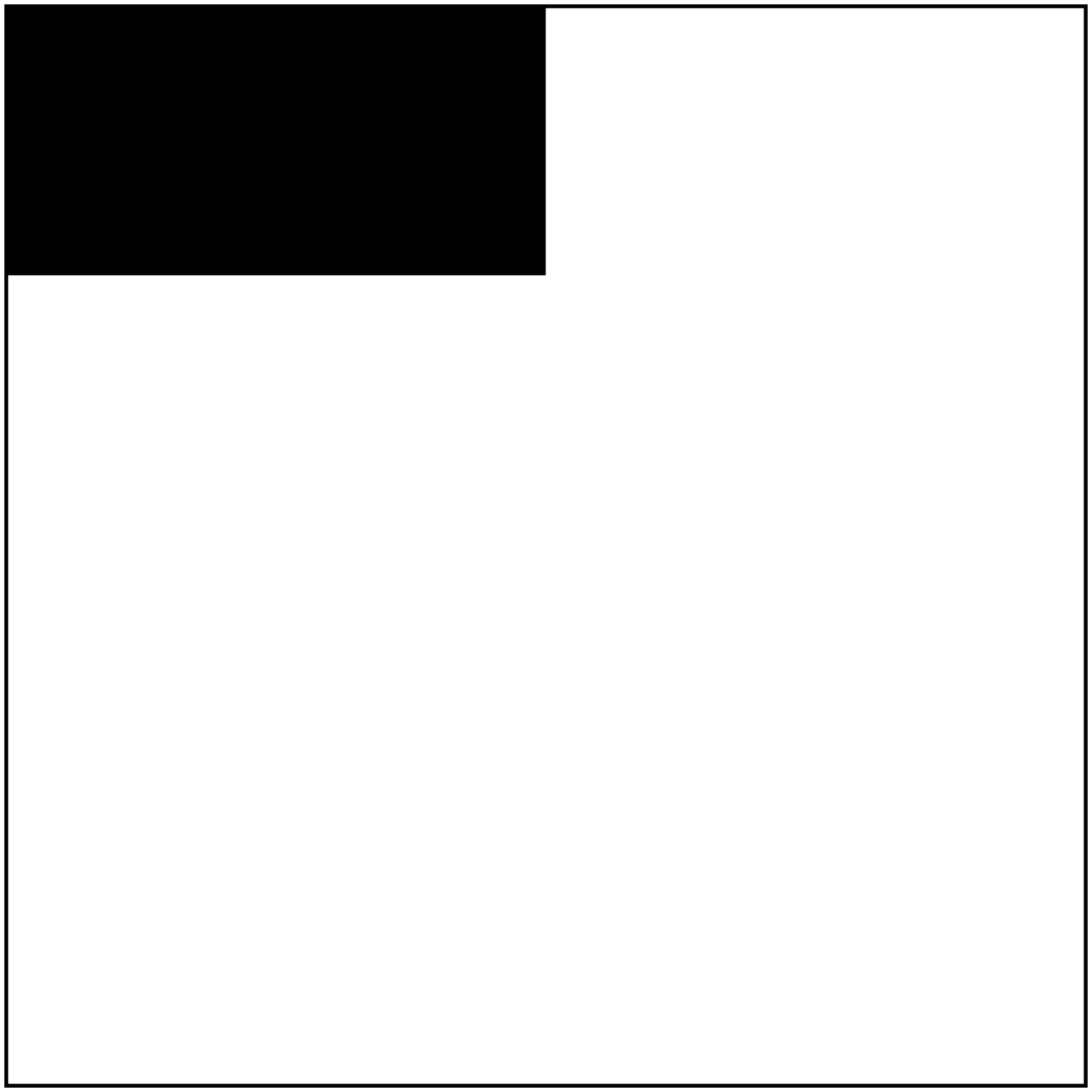}
\includegraphics[width=3cm]{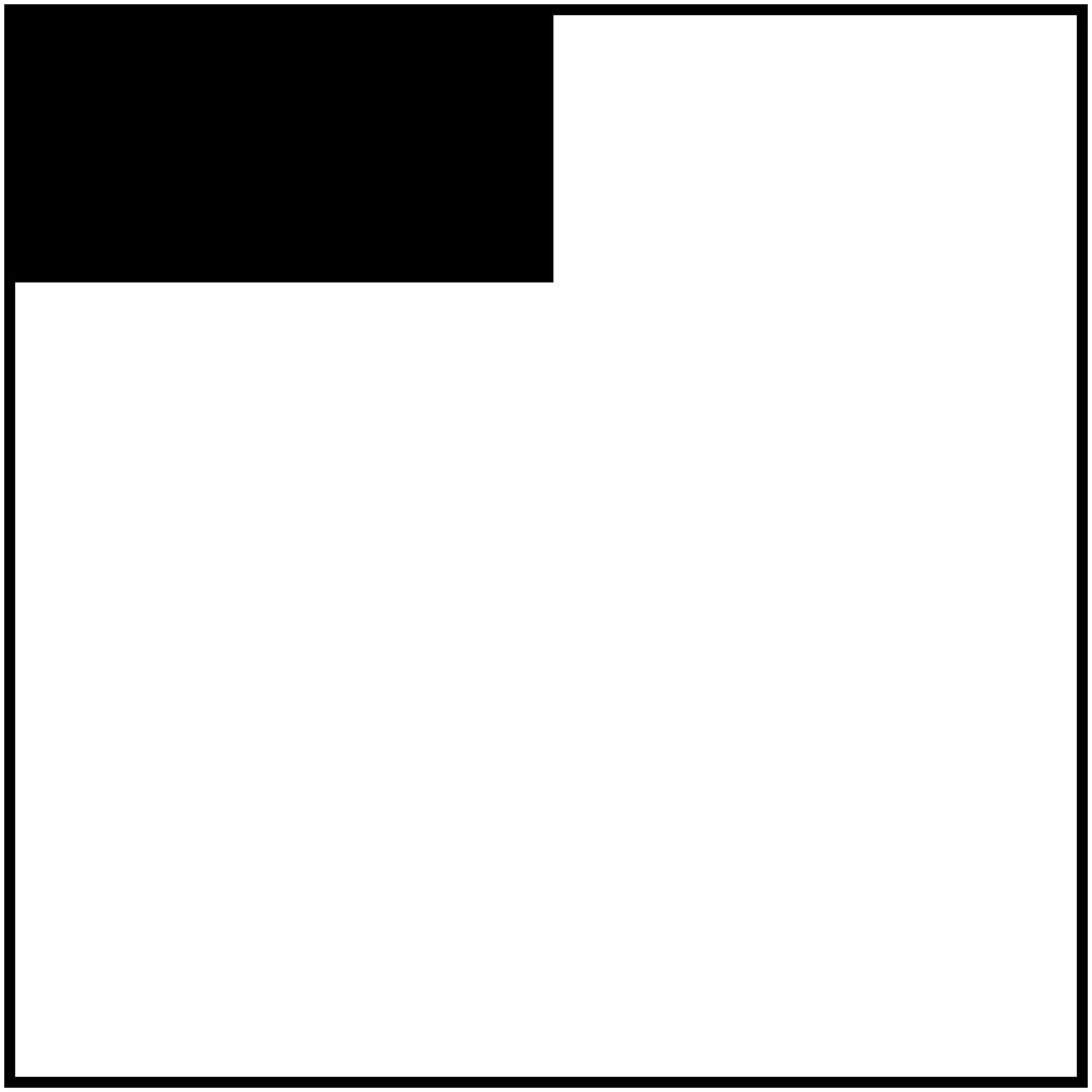}
\includegraphics[width=3cm]{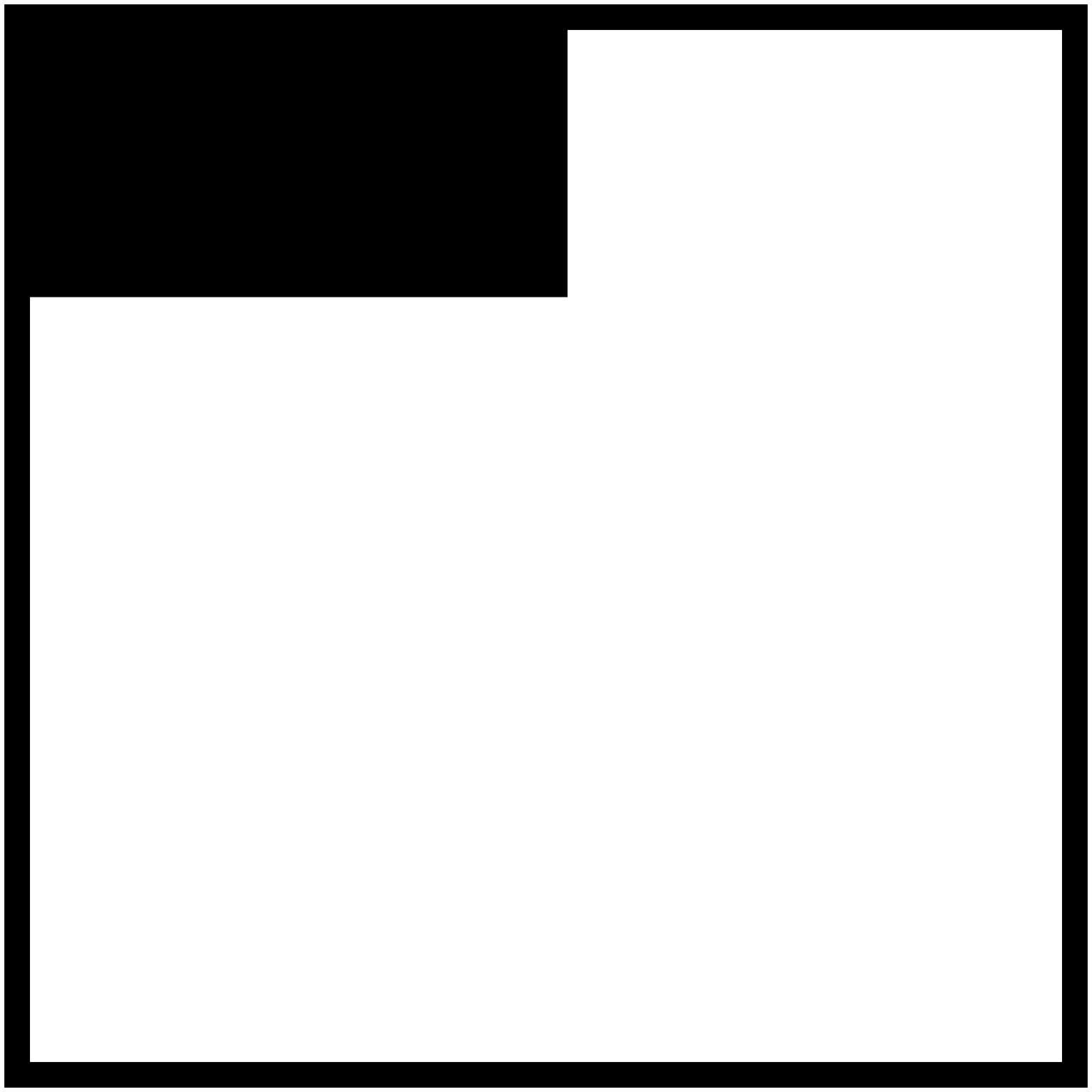}
\includegraphics[width=3cm]{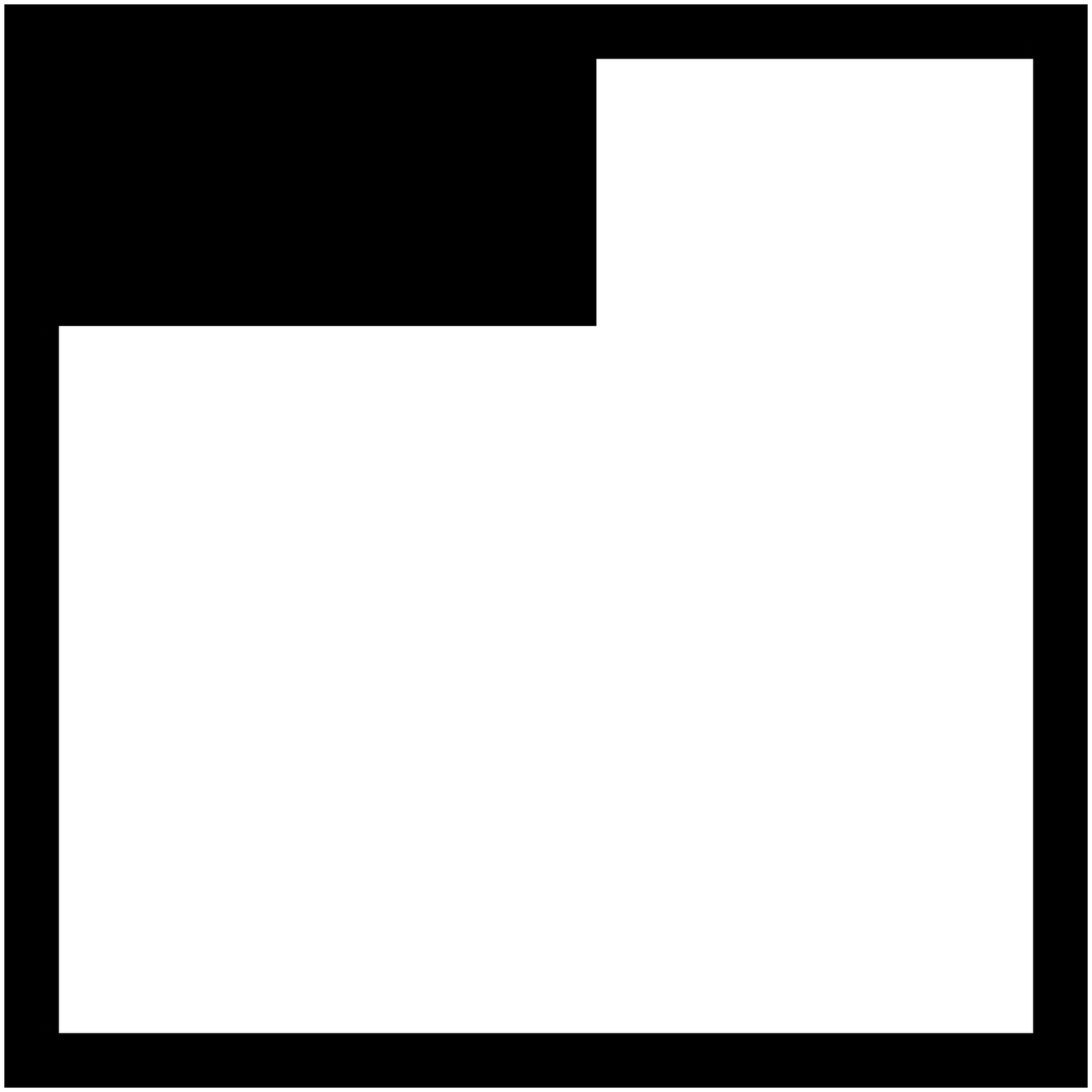}
\includegraphics[width=3cm]{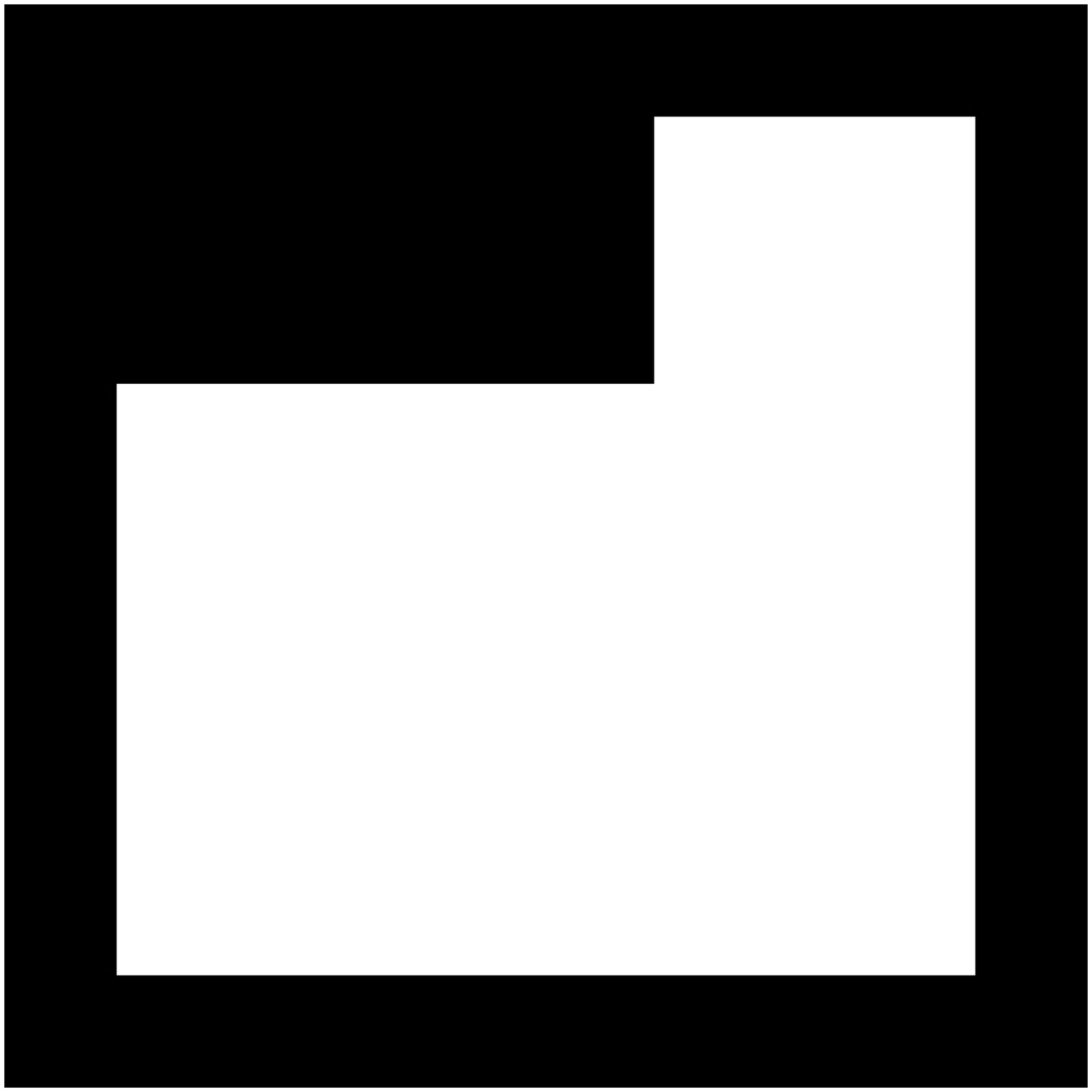}
\includegraphics[width=3cm]{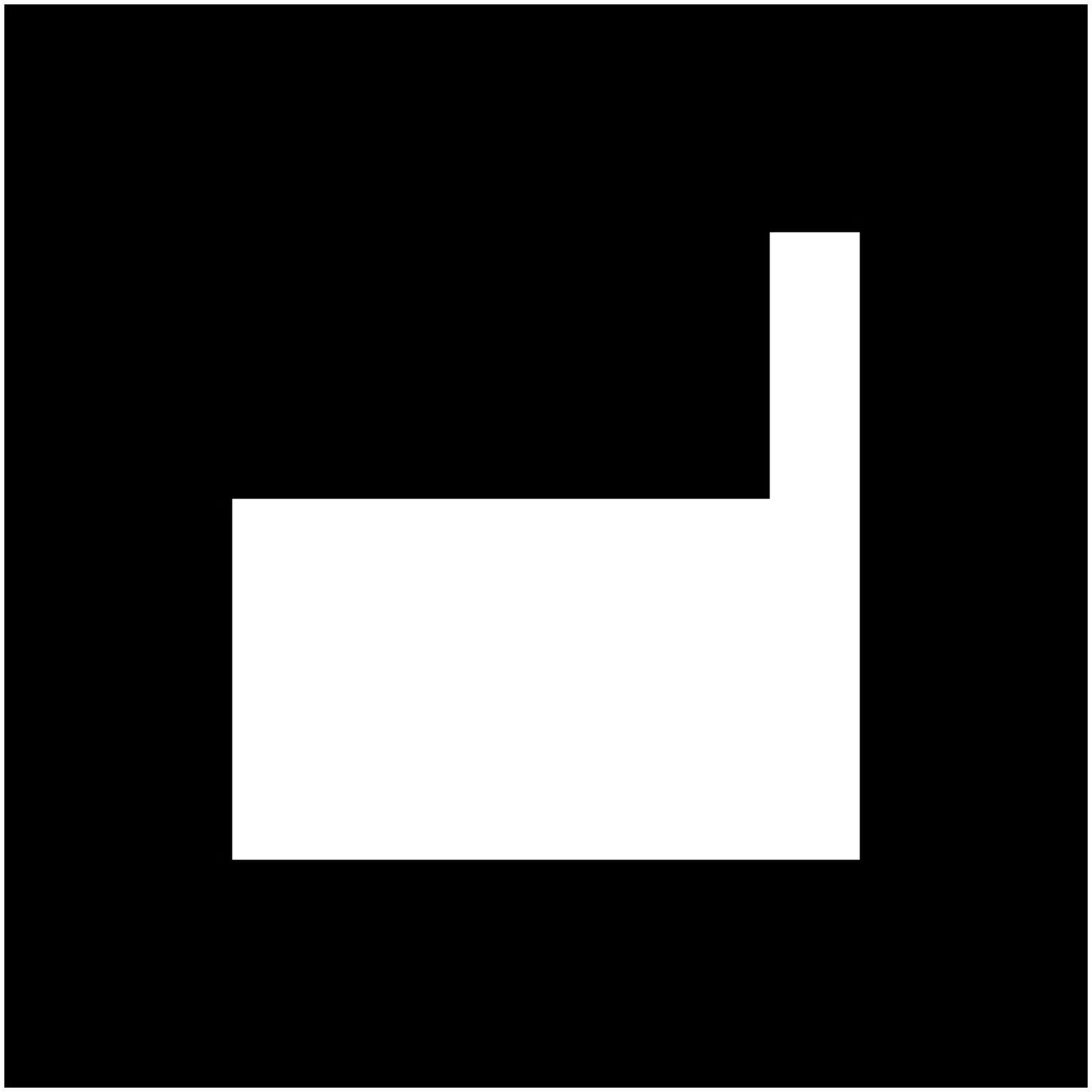}
\caption{Mask used to simulate a gap in the data \emph{(top left)}, 
and the modified masks at scales 1 \emph{(top right)} through 5 \emph{(bottom left)}. The discarded pixels
are in black. }\label{Masks}
\end{center}
\end{figure}

Finally, in order to investigate the benefits of using wSMICA in place of SMICA when gaps are inserted in the data, the mask shown on figure \ref{Masks} was 
applied onto the mixture maps. The case where no data is missing was also considered for the sake of comparison. 
In each of these two particular configurations, spectral matching was assessed and optimized both at the output of the five wavelet filters $\psi_1,\ldots, \psi_5$ associated
to higher frequency details, and on the corresponding five bands in Fourier space, as shown on figure \ref{Filtres2}. This latter choice of frequency bands 
is simply made to ease comparison between SMICA and wSMICA. It may be argued that this choice is probably not optimal
to run SMICA. But, in fact, the optimal selection of filters is clearly a meaningful question  both for SMICA and wSMICA. This will require further investigation.\\
 
\subsection{Preliminary results}

Preliminary experiments were conducted in the case of vanishing instrumental noise
variance, with a square $3 \times 3$ mixing matrix. It was mentioned before
that in this limit, the spectral matching objective boils down to the joint 
diagonalization of covariance matrices. Further, taking the mixing matrix
to be the identity matrix (\emph{i.e.} try to separate sources which are not actually mixed ),
it is possible to gain some insight on the spectral diversity of the independent sources, for a given choice of bands or filters. 
Indeed, the performance of the independent component separation methods based
on spectral matching depend highly on spectral diversity.
\\\\
The following steps were repeated 1000 times:\\
\begin{itemize}   
\item[$\bullet$]{ randomly pick one of each component maps out of the available 200 CMB maps, 30 dust maps and 1500 SZ maps. }
\item[$\bullet$]{ calculate covariance matrices in the five wavelet or Fourier bands, both with and without 
masking part of the maps, as is  all described above.}
\item [$\bullet$]{ normalize each source so that its total energy over the five bands is equal to one.}
\item[$\bullet$]{use the algorithm in \cite{Pha2} to jointly diagonalize the covariances in each configuration, and keep 
the resulting separating matrices.}\\
\end{itemize}

If the sources have satisfactory spectral properties, the obtained separating matrices should not 
depart drastically from the identity matrix. Moreover, denoting $\mathcal{A}$ any invertible $3 \times 3$ mixing matrix,
and $\widehat{\mathcal{A}}^{-1} $ the resulting separating matrix, it is shown in \cite{Car2} that the variances 
of the off-diagonal terms in $\widehat{ \mathcal{A}}^{-1} \mathcal{A} $ depend only on spectral diversity, in the case of 
Gaussian sources. In fact, to assess the effect of any non-Gaussianity or non-stationarity in the source templates, the same 
experiment was repeated on Gaussian maps generated with the same spectra as the CMB, Dust and SZ 
components. In any case, the independent source components are separated using:
\begin{equation}
\widehat{S} = \widehat{ \mathcal{A}}^{-1} \mathcal{A}  S = \mathcal{I}S
\label{SEP1}
\end{equation}
so that with the above normalization, the square of any off-diagonal term $\mathcal{I}_{ij}$ is directly related 
to the residual level of component $j$ in the recovered component $i$. \\

\begin{figure}[!h]
\begin{center}
\includegraphics[width=8cm]{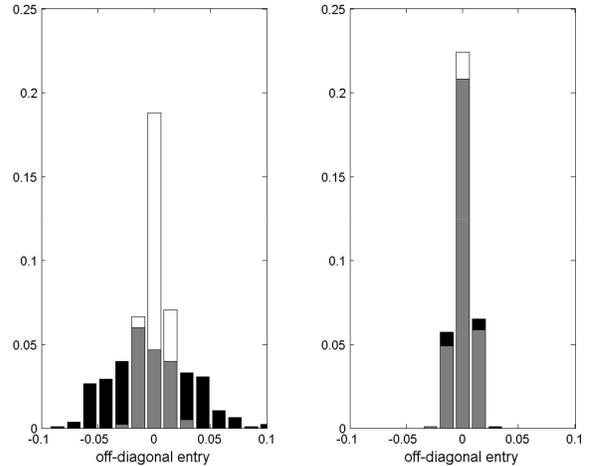} 
\caption{Histograms of the off diagonal term corresponding to the residual corruption of "CMB" by "Dust"
while separating Gaussian maps generated with the same power spectra as the astrophysical components, 
by joint diagonalization of covariance matrices in Fourier \emph{(left)} and wavelet \emph{(right)} space,
with \emph{(black, which appears grey when seen through white )} and without \emph{(white)} masking part of the the data. 
The dark widest histogram on the left highlights the impact of masking on source separation based on Fourier covariances. }\label{histo_cmb}
\end{center}
\end{figure}

\begin{table}[!h]
\begin{center}
\footnotesize{
\begin{tabular}{@{} |c|c|c|c|@{}}
\hline
&  & & \\
& $NM$ & $M$& $Han$ \\
&  & & \\
\hline
&  &  &\\
$\mathcal{I}_{1,2}$ & $0.097\quad\mathit{\underline{0.0076}}$ & $0.074\quad\underline{\mathit{0.038}}$  & $0.024$ \\
&  &  &\\
\hline
&  & & \\
$\mathcal{I}_{1,3}$ & $0.0049\quad\mathit{0.0044 }$ & $0.005\quad\mathit{0.006 }$  & $0.0094$\\
&  & & \\
\hline
&  &  &\\
$\mathcal{I}_{2,1}$ & $0.017\quad\mathit{0.0066 }$ & $0.018\quad\mathit{0.01 }$  & $0.017$\\
&  &  &\\
\hline
&  &  &\\
$\mathcal{I}_{2,3}$ & $0.0064\quad\mathit{0.0077 }$ & $0.0066\quad\mathit{0.0096 }$  & $0.011$\\
&  &  &\\
\hline
&  &  &\\
$\mathcal{I}_{3,1}$ & $0.0024\quad\mathit{0.0026 }$ & $0.0028\quad\mathit{0.0037 }$  & $0.0039$\\
&  &  &\\
\hline
&  &  &\\
$\mathcal{I}_{3,2}$ & $0.0054\quad\mathit{0.0071 }$  & $0.0054\quad\mathit{0.0079 }$ & $0.01$ \\
&  &  &\\
\hline
\end{tabular}}
\caption{Standard deviations of the off-diagonal entries $\mathcal{I}_{ij}$ defined by (\ref{SEP1}) obtained while separating realistic
component maps by joint diagonalization of covariance matrices in \textbf{Fourier} space,  with ($M$) or without masking ($NM$) part 
of the data, or applying an apodizing Hanning window ($Han$). Components 1, 2 and 3 respectively stand for CMB, Dust and SZ. The numbers in \emph{italic} were obtained with Gaussian maps and the underlined
numbers correspond to the histograms in figure \ref{histo_cmb}. }\label{RES2}
\end{center}
\end{table}

\begin{table}[!h]
\begin{center}
\footnotesize{
\begin{tabular}{@{} |c|c|c| @{}}
\hline
&  & \\
& $NM$ & $M$  \\
&  &  \\
\hline
&  &  \\
$\mathcal{I}_{1,2}$ & $0.015\quad \underline{\mathit{0.0071 }}$ & $0.018\quad\underline{\mathit{0.0079 }}$ \\
&  & \\
\hline
&  &  \\
$\mathcal{I}_{1,3}$ & $0.0025\quad\mathit{0.0029 }$ & $0.0028\quad\mathit{0.0031 }$ \\
&  &  \\
\hline
&  &  \\
$\mathcal{I}_{2,1}$ & $0.016\quad\mathit{0.0077 }$ & $0.019\quad\mathit{0.0089 }$ \\
&  &  \\
\hline
&  &  \\
$\mathcal{I}_{2,3}$ & $0.0041\quad\mathit{0.0051 }$ & $0.0048\quad\mathit{0.0075 }$ \\
&  &  \\
\hline
&  &  \\
$\mathcal{I}_{3,1}$ & $0.0024\quad\mathit{0.0029 }$ & $0.003\quad\mathit{0.0039 }$ \\
&  &  \\
\hline
&  &  \\
$\mathcal{I}_{3,2}$ & $0.0039\quad\mathit{0.0054 }$  & $0.0053\quad\mathit{0.0085 }$ \\
&  &  \\
\hline
\end{tabular}}
\caption{Standard deviations of the off-diagonal entries $\mathcal{I}_{ij}$ defined by (\ref{SEP1}) obtained while separating realistic
component maps by joint diagonalization of covariance matrices in \textbf{wavelet} space,  with ($M$) and without masking ($NM$) part 
of the data. Components 1, 2 and 3 respectively stand for CMB, Dust and SZ. The numbers in \emph{italic} were obtained with Gaussian maps and the underlined
numbers correspond to the histograms in figure \ref{histo_cmb}. }\label{RES3}
\end{center}
\end{table}

The histograms on figure \ref{histo_cmb} are for the off diagonal term corresponding to the residual corruption of CMB by \emph{Gaussian} Dust
in the second set of experiments. In tables \ref{RES2} and \ref{RES3}, the results obtained with
the synthetic component maps are given as well as those obtained with the Gaussian maps, in terms of the standard deviations of the 
off-diagonal entries $\mathcal{I}_{ij}$ defined by (\ref{SEP1}).\\

Interestingly, when working on Gaussian maps without masks, using covariances in Fourier space 
or in wavelet space gives similar performances. It is also satisfactory, when covariances in wavelet 
space are used with Gaussian maps, that each computed standard deviation only slightly increases when a mask 
is applied on the data. Indeed, as a consequence of incomplete coverage, there are
less samples from which to estimate the covariances.
This increase is also observed when covariances in Fourier space are used with the Gaussian maps 
but it can be as high as five-fold and it does not affect all coefficients the same way. Although this 
can again  be attributed to the reduced data size, the lowered spectral diversity between components,
because of the correlations and smoothing induced in Fourier space by the mask, is also part of the explanation. In fact, as 
shown on figure \ref{mixtures1}, CMB and dust spatial power spectra are somewhat similar, \emph{i.e.} show low 
spectral diversity, and further smoothing can only degrade the performance of the
source separation algorithm based on Fourier covariances.\\  

In the case of realistic component maps, we note first that the comparison of the performance of component separation
using wavelet covariances with and without mask again agrees with the different data sizes, which is not the case
with covariances in Fourier space. Next, whether covariances in Fourier or wavelet space are used, we note that 
the terms coupling CMB and Dust are again much higher in magnitude, even on complete maps.
It seems that the actual
non-stationarity and non-Gaussianity of the realistic component maps are relevant issues. Another point is 
that the CMB and Dust templates as 
in figure \ref{Templates} exhibit sharp edges compared to SZ and this 
inevitably disturbs spectral estimation using a simple DFT. To assess
this effect, simulations were also conducted where the covariances in Fourier space were computed after
an apodizing Hanning window was applied on the complete data maps. The results
 reported in table \ref{RES2}, to be compared to table \ref{RES3}, do indicate a slightly positive effect
of windowing, but still the separation using wavelet covariances appears
better.\\\\

\subsection{Realistic experiments}

The above preliminary results clearly point out in the noiseless case the advantageous use of wavelets to easily escape
the very bad impact that gaps and sharp edges actually have on the performance of the source separation using covariances 
in Fourier space. Hence this is strong encouragement to move on to investigating the effect of additive noise on the mixture
 maps according to (\ref{model1}), using SMICA and its extension wSMICA. 
We note that although in the case of wSMICA the link with maximum likelihood is not 
as strongly asserted as with SMICA, the optimization algorithm used in the simulations hereafter consists in both cases of 
the same heuristic succession of EM and BFGS steps and initialization is done as discussed in paragraph \ref{OPTIM}.\\
 
Picking at random one of each component maps out of the available 200 CMB maps, 30 dust maps and 1500 SZ maps, $1000$ synthetic
mixture maps were generated as previously described, for each of the 5 noise levels chosen. Then, component separation was conducted using 
the spectral matching algorithms SMICA and wSMICA both with and without part of the maps being masked.
Now, each run of SMICA and wSMICA on
 the data returns estimates $\widehat{A}_f$ and $\widehat{A}_w$ of the mixing matrix. Clearly, these estimates are subject
to the indeterminacies inherent to the instantaneous linear mixture model (\ref{model1}). Indeed, in the case 
where optimization is over all parameters $\theta$,
it is obvious that any simultaneous permutation of the columns of $A$ and of the lines of $S$ leaves the model unchanged. The same occurs when
exchanging a scalar possibly negative factor between any column in $A$ and the corresponding line in $S$. Therefore, columnwise comparison of 
$\widehat{A}_f$ and $\widehat{A}_w$ to the original mixing matrix $A$ requires first fixing these indeterminacies. This is done by hand after $\widehat{A}_f$ and $\widehat{A}_w$
have been normalized columnwise.\\ 
 
\begin{figure}[!h]
\begin{center}
\includegraphics[width=7cm]{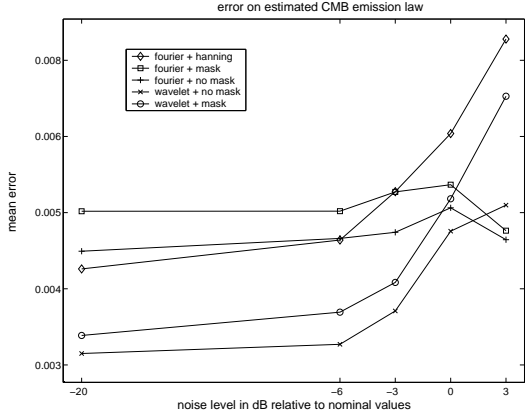} 
\caption{Comparison of the mean squared errors on the estimation of the emissivity of \textbf{CMB} as a function of noise 
in five different configurations namely : wSMICA without mask, wSMICA with mask, fSMICA without mask, fSMICA with mask, 
fSMICA with Hanning apodizing window.}\label{cmb_emis}
\end{center}
\end{figure}

\begin{figure}[!h]
\begin{center}
\includegraphics[width=7cm]{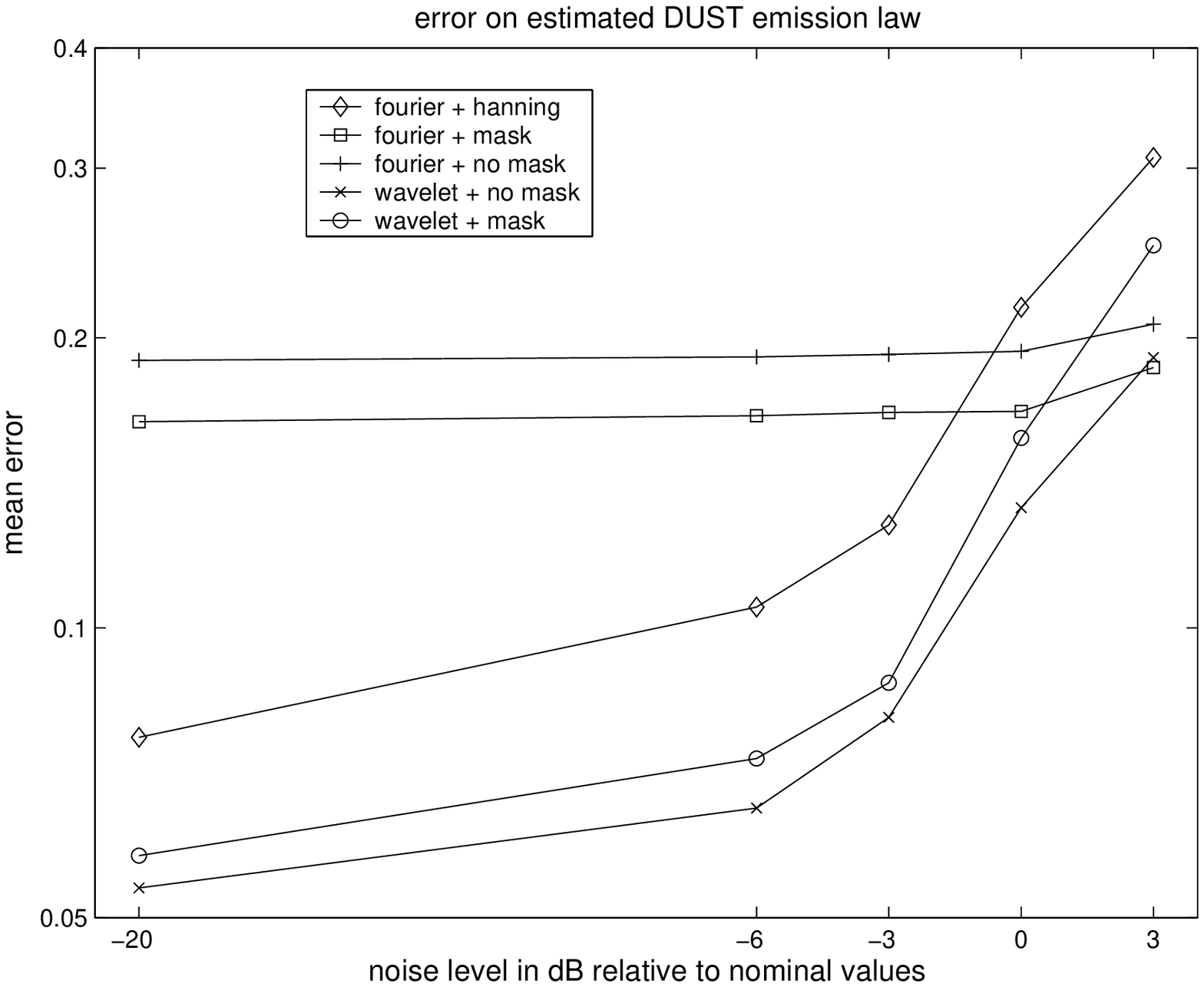} 
\caption{Comparison of the mean squared errors on the estimation of the emissivity of \textbf{DUST} as a function of noise 
in five different configurations namely : wSMICA without mask, wSMICA with mask, fSMICA without mask, fSMICA with mask, 
fSMICA with Hanning apodizing window.}\label{dust_emis}
\end{center}
\end{figure}

\begin{figure}[!h]
\begin{center}
\includegraphics[width=7cm]{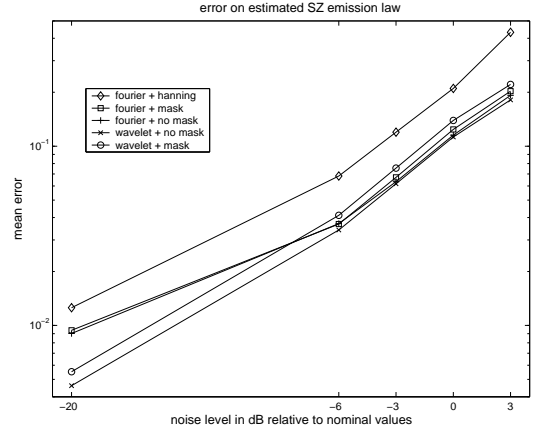} 
\caption{Comparison of the mean squared errors on the estimation of the emissivity of \textbf{SZ} as a function of noise 
in five different configurations namely : wSMICA without mask, wSMICA with mask, fSMICA without mask, fSMICA with mask, 
fSMICA with Hanning apodizing window.}\label{sz_emis}
\end{center}
\end{figure}

The results we report next concentrate on the statistical properties of $\widehat{A}_f$ and $\widehat{A}_w$ as estimated from the 1000 runs of the 
two competing methods in the several configurations retained. In fact, the correct estimation of the mixing matrix in model
(\ref{model1}) is a relevant issue for instance when it comes to dealing  with the
cross calibration of the different detectors. Figures \ref{cmb_emis}, \ref{dust_emis} and \ref{sz_emis} show the results
obtained,  using the quadratic norm
\begin{equation}
QE_j =\left(  \sum_{i = 1}^m \left( A_{ij} - \widehat{A}_{ij} \right)^2 \right)^\frac{1}{2}
\end{equation}
with $\widehat{A} = \widehat{A}_f\textrm{ or } \widehat{A}_w$  and $j = \textrm{CMB, DUST or SZ}$, to assess the residual errors on the estimated 
emissivities of each component. The plotted curves show how the mean of the above positive error measure varies with increasing noise 
variance. For the particular case of CMB, table \ref{RES4} gives the estimated standard deviations of 
the relative errors 
\begin{equation}
\frac{A_{ij} - \widehat{A}_{ij} }{ A_{ij} }
\end{equation}
on the estimated CMB emissivity in the six channels of Planck's HFI in the different 
configurations retained.\\

\begin{table}[!h]
\begin{center}
\tiny{
\begin{tabular}{@{} |c|c|c|c|c|c| @{}}
\hline
&  & &  & & \\
& WNM & WM &  FNM & FM & FHan \\
&  & &  & & \\
\hline
&  & &  & & \\
 & $ 4.4\!*\!10^{-4} $&$  5.0\!*\!10^{-4}  $&$ 6.2\!*\!10^{-4}  $&$ 7.3\!*\!10^{-4} $&$  7.2\!*\!10^{-4}$\\
 & $ 5.4\!*\!10^{-4} $&$  7.5\!*\!10^{-4}  $&$ 7.1\!*\!10^{-4}  $&$ 8.5\!*\!10^{-4} $&$  9.5\!*\!10^{-4}$\\
$\mathcal{A}_{11}$ & $ 6.6\!*\!10^{-4} $&$  9.2\!*\!10^{-4}  $&$ 8.2\!*\!10^{-4}  $&$ 8.9\!*\!10^{-4} $&$  1.3\!*\!10^{-3} $\\
 & $ 9.4\!*\!10^{-4} $&$  1.2\!*\!10^{-3}  $&$ 1.0\!*\!10^{-3}  $&$ 1.0\!*\!10^{-3}  $&$  1.7\!*\!10^{-3} $\\
 & $ 1.2\!*\!10^{-3} $&$  1.7\!*\!10^{-3}  $&$ 1.2\!*\!10^{-3}  $&$ 1.4\!*\!10^{-3}  $&$  2.3\!*\!10^{-3} $\\
&  & &  & & \\
\hline 
&  & &  & & \\
  & $ 1.6\!*\!10^{-4}  $&$ 2.1\!*\!10^{-4}  $&$ 2.1\!*\!10^{-4}  $&$ 2.0\!*\!10^{-4}  $&$ 2.7\!*\!10^{-4}$\\
  & $ 5.3\!*\!10^{-4}  $&$ 7.8\!*\!10^{-4}  $&$ 5.6\!*\!10^{-4}  $&$ 5.7\!*\!10^{-4}  $&$ 1.0\!*\!10^{-3}$\\
$\mathcal{A}_{21}$  & $ 7.0\!*\!10^{-4}  $&$ 1.1\!*\!10^{-3}  $&$ 7.6\!*\!10^{-4}  $&$ 8.4\!*\!10^{-4}  $&$ 1.4\!*\!10^{-3}$\\
  & $ 1.0\!*\!10^{-3}  $&$ 1.6\!*\!10^{-3}  $&$ 1.0\!*\!10^{-3}  $&$ 1.0\!*\!10^{-3}  $&$ 2.1\!*\!10^{-3}$\\
  & $ 1.4\!*\!10^{-3}  $&$ 2.2\!*\!10^{-3}  $&$ 1.5\!*\!10^{-3}  $&$ 1.7\!*\!10^{-3}  $&$ 3.1\!*\!10^{-3}$\\
&  & &  & & \\
\hline 
&  & &  & & \\
  & $1.5\!*\!10^{-3} $&$  1.8\!*\!10^{-3}  $&$ 2.2\!*\!10^{-3}  $&$ 2.5\!*\!10^{-3}  $&$ 2.3\!*\!10^{-3}$\\
 & $ 1.7\!*\!10^{-3} $&$  2.1\!*\!10^{-3}  $&$ 2.3\!*\!10^{-3}  $&$ 2.6\!*\!10^{-3}  $&$ 2.9\!*\!10^{-3}$\\
 $\mathcal{A}_{31}$ & $ 2.1\!*\!10^{-3} $&$  2.6\!*\!10^{-3}  $&$ 2.6\!*\!10^{-3}  $&$ 2.8\!*\!10^{-3}  $&$ 3.7\!*\!10^{-3}$\\
 & $ 2.7\!*\!10^{-3} $&$  3.0\!*\!10^{-3}  $&$ 2.9\!*\!10^{-3}  $&$ 3.0\!*\!10^{-3}  $&$ 4.2\!*\!10^{-3}$\\
 & $ 3.3\!*\!10^{-3} $&$  4.6\!*\!10^{-3}  $&$ 3.3\!*\!10^{-3}  $&$ 3.5\!*\!10^{-3}  $&$ 6.1\!*\!10^{-3}$\\
&  & &  & & \\
\hline 
&  & &  & & \\
 & $ 1.8\!*\!10^{-2} $&$  2.0\!*\!10^{-2}  $&$ 2.7\!*\!10^{-2}  $&$ 3.0\!*\!10^{-2} $&$  2.5\!*\!10^{-2}$\\
 &  $1.9\!*\!10^{-2} $&$  2.1\!*\!10^{-2}  $&$ 2.7\!*\!10^{-2}  $&$ 2.1\!*\!10^{-2} $&$  2.7\!*\!10^{-2}$\\
$\mathcal{A}_{41}$ &  $2.1\!*\!10^{-2} $&$  2.4\!*\!10^{-2}  $&$ 2.8\!*\!10^{-2}  $&$ 3.1\!*\!10^{-2} $&$  2.9\!*\!10^{-2}$\\
 &  $2.7\!*\!10^{-2} $&$  2.8\!*\!10^{-2}  $&$ 3.1\!*\!10^{-2}  $&$ 3.0\!*\!10^{-2} $&$  3.5\!*\!10^{-2}$\\
 &  $3.0\!*\!10^{-2} $&$  4.1\!*\!10^{-2}  $&$ 2.5\!*\!10^{-2}  $&$ 2.7\!*\!10^{-2} $&$  4.9\!*\!10^{-2}$\\
&  & &  & & \\
\hline 
&  & &  & & \\
 & $ 4.0\!*\!10^{-1} $&$  4.5\!*\!10^{-1}  $&$ 6.1\!*\!10^{-1}  $&$ 6.6\!*\!10^{-1}  $&$ 5.6\!*\!10^{-1}$\\
 &  $4.2\!*\!10^{-1} $&$  4.7\!*\!10^{-1}  $&$ 6.1\!*\!10^{-1}  $&$ 6.5\!*\!10^{-1}  $&$ 5.8\!*\!10^{-1}$\\
$\mathcal{A}_{51}$ &  $4.5\!*\!10^{-1} $&$  5.0\!*\!10^{-1}  $&$ 6.1\!*\!10^{-1}  $&$ 6.7\!*\!10^{-1}  $&$ 6.4\!*\!10^{-1}$\\
 &  $5.7\!*\!10^{-1} $&$  5.9\!*\!10^{-1}  $&$ 6.7\!*\!10^{-1}  $&$ 6.7\!*\!10^{-1}  $&$ 7.5\!*\!10^{-1}$\\
 &  $6.2\!*\!10^{-1} $&$  8.4\!*\!10^{-1}  $&$ 5.0\!*\!10^{-1}  $&$ 5.5\!*\!10^{-1}  $&$ 1.0$\\
&  & &  & & \\
\hline 
&  & &  & & \\
 &  $5.7\!*\!10^{1}  $& $ 6.2\!*\!10^{1}  $&$ 8.5\!*\!10^{1} $&$  9.2\!*\!10^{1} $&$  7.8\!*\!10^{1}$\\
 &  $5.8\!*\!10^{1}  $&$ 6.5\!*\!10^{1}  $&$ 8.6\!*\!10^{1} $&$  9.1\!*\!10^{1} $&$  8.1\!*\!10^{1}$\\
$\mathcal{A}_{61}$ & $ 6.2\!*\!10^{1}  $&$ 6.9\!*\!10^{1}  $&$ 8.6\!*\!10^{1} $&$  9.4\!*\!10^{1} $&$  8.9\!*\!10^{1}$\\
 &  $7.9\!*\!10^{1}  $&$ 8.2\!*\!10^{1}  $&$ 9.3\!*\!10^{1} $&$ 9.2\!*\!10^{1} $&$  1.0\!*\!10^{2}$\\
 & $ 8.6\!*\!10^{1}  $&$ 1.2\!*\!10^{2}  $&$ 6.9\!*\!10^{1} $&$  7.7\!*\!10^{1} $&$  1.4\!*\!10^{2}$\\
&  & &  & & \\
\hline 

\end{tabular}}
\caption{\textbf{Standard deviations} of the relative errors on the 
estimated emissivities $\mathcal{A}_{i1}$ of CMB in Planck's HFI six channels. The colunm labels WNM, WM, FNM, FM, FHan are for the different 
configurations, respectiveley: wSMICA without mask, wSMICA with mask, fSMICA without mask, fSMICA with mask, fSMICA with Hanning apodizing window. 
The five figures in each box are for noise variance -20, -6, -3, 0 and 3~dB from nominal Planck values.}\label{RES4}
\end{center}
\end{table}

\begin{figure}[!h]
\begin{center}
\includegraphics[width=7cm]{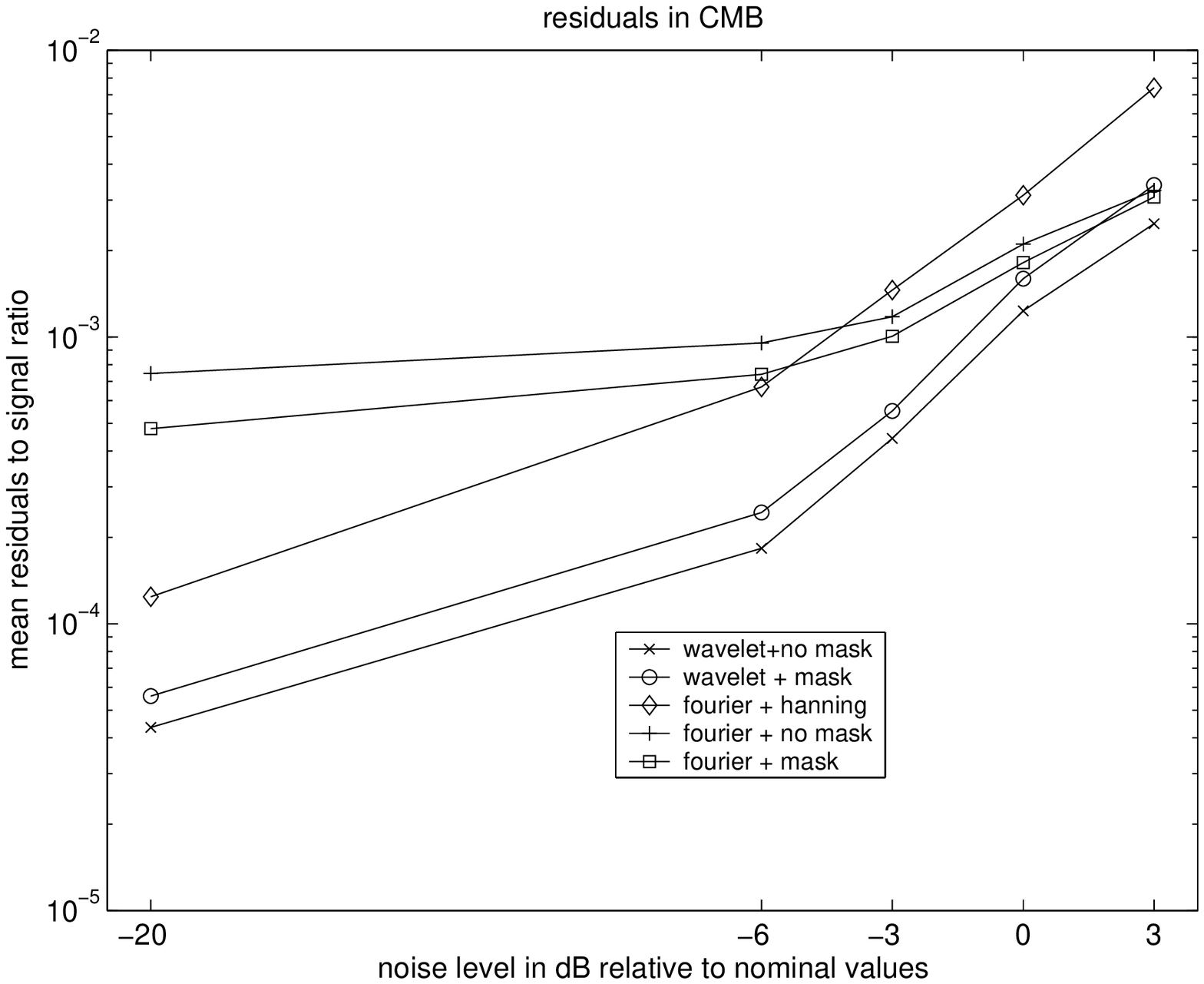} 
\caption{Comparison of the mean ISR for \textbf{CMB} as a function of noise 
in five different configurations namely : wSMICA without mask, wSMICA with mask, fSMICA without mask, fSMICA with mask, 
fSMICA with Hanning apodizing window.}\label{CMB_res}
\end{center}
\end{figure}

\begin{figure}[!h]
\begin{center}
\includegraphics[width=7cm]{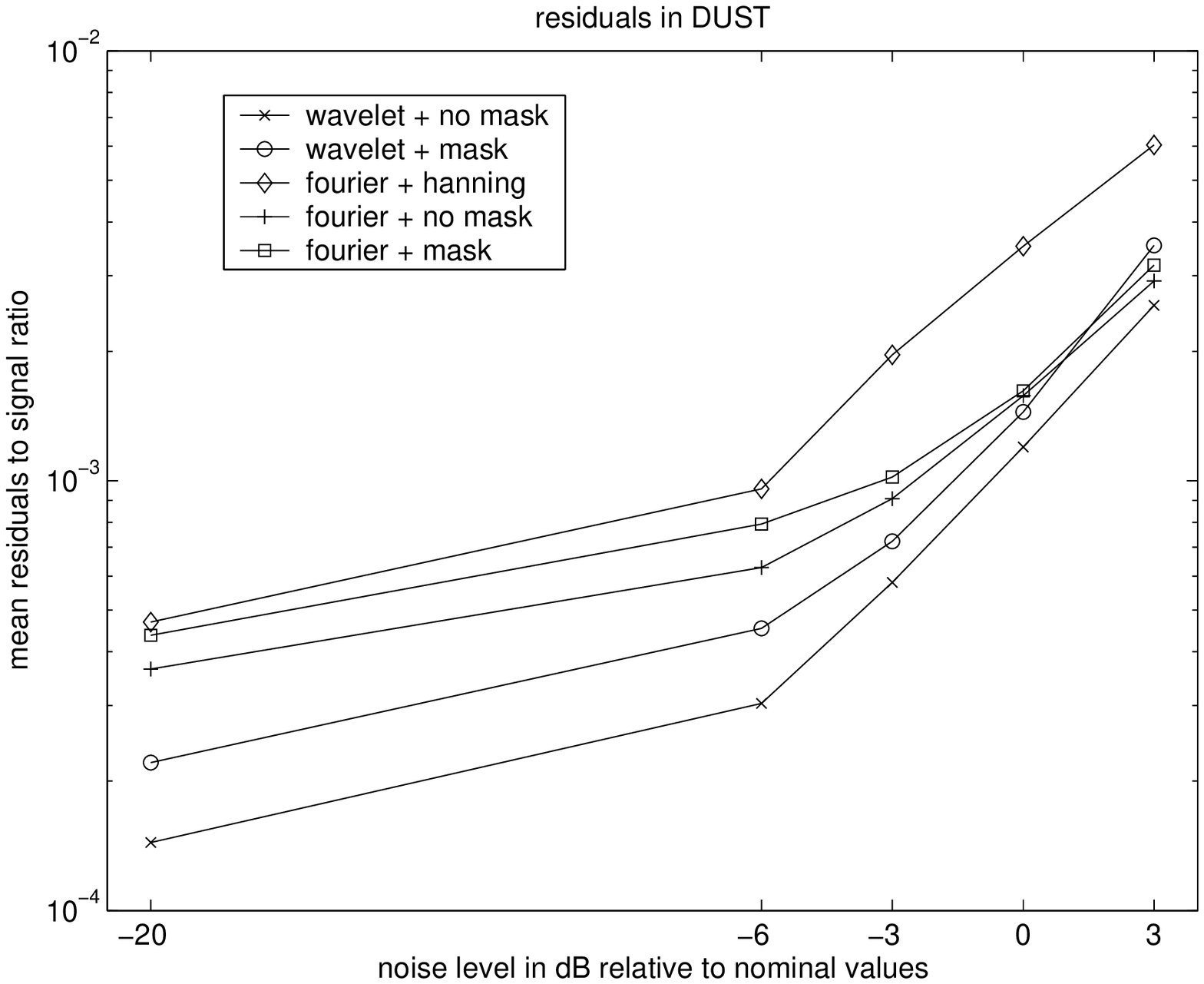} 
\caption{Comparison of the mean ISR for \textbf{DUST} as a function of noise 
in five different configurations namely : wSMICA without mask, wSMICA with mask, fSMICA without mask, fSMICA with mask, 
fSMICA with Hanning apodizing window.}\label{Dust_res}
\end{center}
\end{figure}

\begin{figure}[!h]
\begin{center}
\includegraphics[width=7cm]{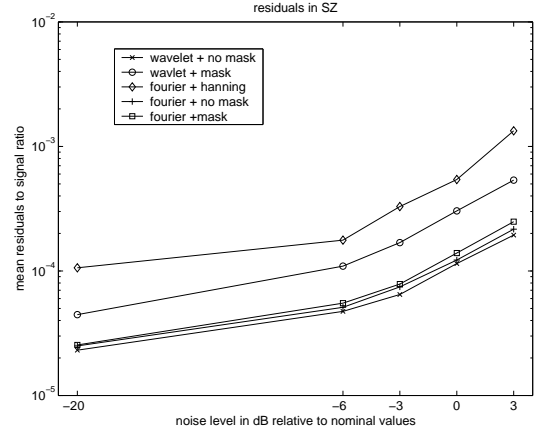} 
\caption{Comparison of the mean ISR for \textbf{SZ} as a function of noise 
in five different configurations namely : wSMICA without mask, wSMICA with mask, fSMICA without mask, fSMICA with mask, 
fSMICA with Hanning apodizing window.}\label{SZ_res}
\end{center}
\end{figure}

Closer to our source separation objective, a more significant way of assessing the quality 
of $\widehat{A}_f$ and $\widehat{A}_w$ as estimators of the mixing matrix $A$, would be to use the following 
signal to interference ratio:
\begin{equation}
ISR_j =  \frac{ \mathcal{I}_{j,j} ^2 \sigma_j^2   }{ \sum_{i \neq j} \mathcal{I}_{j,i} ^2 \sigma_i^2 }  
\end{equation}
where the $\sigma_j$ are the source variances and 
\begin{equation}
\mathcal{I} =  (\widehat{A}\adj \widehat{R}_{N}^{-1} \widehat{A})\inv \widehat{A}\adj \widehat{R}_{N}^{-1} A 
\end{equation}
with $R_N$ the noise covariance. The plots on figures \ref{CMB_res}, \ref{Dust_res} and \ref{SZ_res} show how the mean ISR 
from the 1000 runs of SMICA and wSMICA in different configurations,  varies with increasing noise. 

We note again that the performance of wSMICA behaves as expected when noise increases and if part of the data is missing. 
However this is not always the case with SMICA. Finally this  set of simulations, conducted in a more realistic setting with respect to ESA's Planck
mission, again confirms the higher performance, over Fourier analysis, that we indeed expected  from the use of
wavelets. The latter are able to correctly grab the  spectral content of partly masked data maps and from 
there allow for better component separation.

\section{Conclusion}\label{conclusion}

This paper has presented an extension of the Spectral Matching ICA algorithm to the case where the collected data is both correlated and non stationary, considering maps with gaps as a particular instance of practical significance. It was shown that simply substituting covariance matching in Fourier space by covariance matching in wavelet space enables to cope in the most general and straightforward way with gaps of possibly any shape. Mainly, it is the FIR nature of the wavelet filters used that allows the impact of edges and gaps on the estimated covariances and hence on component separation to be lowered. Optimally choosing the FIR filter-bank regarding a particular application is a possible further enhancement.\\

Results obtained  with simulated astrophysical data as expected from the Planck mission were given and these confirm the benefits of correctly processing existing gaps. Clearly, other possible types of  non-stationarities in the collected data such as spatially varying noise or component variance, etc. can be dealt with very simply in a similar fashion using the wavelet extension of SMICA.\\

In the CMB application, the mixed components have quite different statistical properties : some are expected to be very close to Gaussian whereas others are strongly non Gaussian. Standard ICA methods exploit the non Gaussianity of the mixed components. However, it is not clear yet how best to combine non Gaussianity and spectral diversity in order to perform better source separation. Other features of wavelets which are known to be powerful tools for 
the analysis and sparse representation of structured data might reveal useful here. 

\appendix
\section{Appendix : EM algorithm with constraints on the mixing matrix}\label{annexe1}

Considering $Q$ separate frequency bands of size $n_q$ with $\sum n_q = 1$, the EM functional derived for the instantaneous mixing model 
(\ref{model1}) with independent Gaussian stationary sources $S$ and noise $N$ is:
\begin{equation}
  \Phi(\underline \theta, \theta) = \mathcal{E} \left\{ \log p(X,S|\underline \theta) | \theta    \right\} 
\end{equation}
with $\theta=(A, R_{S,1}, \ldots, R_{S,Q}, R_{N,1},\ldots , R_{N,Q}) $ and
$\underline \theta=(\underline A, \underline R_{S,1} , \ldots, \underline
R_{S,Q}, \underline R_{N,1},\ldots ,\underline R_{N,Q}) $. The maximization step of the EM algorithm seeks then to
maximize $\Phi(\underline \theta, \theta)$ with respect to $\underline \theta$ and the optimal $\underline \theta$ 
is used as the value for $\theta$ at the next EM step, and so on until satisfactory convergence is reached. 
Explicit expressions are easily derived for the optimal $\underline \theta$ in the white noise case 
where an interesting decoupling occurs between the re-estimating equations for 
noise variances, source variances and  the mixing matrix \cite{Car}.\\

\subsection*{Linear equality constraints }

When $A$ is subject to linear constraints, the joint maximization of the EM functional with respect to all model parameters is no
longer easily achieved in general. In fact, one cannot simply decouple the re-estimating rules for the noise parameters and the
mixing matrix and these have to be optimized separately. We give next the modified re-estimating equations for the mixing matrix and 
the source variances in the case of constant noise (\emph{i.e.} $\theta=(A, R_{S,1}, \ldots, R_{S,Q})$ ). \\
 
First, let us exhibit the quadratic dependence of the EM functional $\Phi(\underline\theta , \theta)$ on
$\underline A$ :
\begin{multline} \label{EM1}
\Phi(\underline\theta , \theta) = -\frac{1}{2} \sum_q n_q  \trace \Big( \underline{A} R_q^{ss} \underline{A}\adj R_{N,q}\inv \\
 -\underline{A} R_q^{xs\dagger} R_{N,q}\inv - R_q^{xs} \underline{A}\adj  R_{N,q}\inv \Big) + const_{\underline A}
\end{multline}
where
\begin{align}
C_q &=& (A\adj R_{N,q} \inv A + R_{S,q}\inv )\inv \\
W_q &=& (A\adj R_{N,q} \inv A + R_{S,q}\inv )\inv  A\adj R_{N,q}\inv \\
R_q^{xs} &=& \widehat R_{X,q} W_q\adj \\
R_q^{ss} &=& W_q \widehat R_{X,q} W_q\adj + C_q  
\end{align}

\textbf{In the white noise case}, $R_{N,q}=R_N$, equation (\ref{EM1}) becomes:
\begin{multline}
\Phi(\underline\theta , \theta) = -\frac{1}{2} \trace \Big( (\underline{A} - R^{xs} R^{ss-1}  )R^{ss}\\
(\underline{A} - R^{xs}R^{ss-1} )^\dagger R_N\inv\Big) + const_{\underline A}
\end{multline}
where : 
\begin{equation}
R^{xs}=\sum_q n_q R_q^{xs} \quad \textrm{and} \quad R^{ss} = \sum_q n_q R_q^{ss}
\end{equation}
Again, this can be re-written as : 
\begin{equation}
\Phi(\underline\theta , \theta) = -\frac{1}{2}  ( \underline{\mathcal{A}} - \mathcal{M} ) 
\mathcal{Q}(\underline{\mathcal{A}} - \mathcal{M} )\adj + const_{\underline A}
\end{equation}
where:
\begin{equation}
\underline{\mathcal{A}} = \vect \underline{A} \quad \textrm{,} \quad \mathcal{Q} = \underline{R_N}\inv \otimes \sum_q n_q R_q^{ss}
\end{equation}
\begin{equation}
\mathcal{M} = \vect \left(\left( \sum_q n_q R_q^{ys}\right) \left(\sum_q n_q R_q^{ss} \right)\inv  \right)
\end{equation}
With ``$\vect$'', we build a column vector with the entries of a matrix taken along its lines. 
Now let us consider linear constraints on the mixing matrix, specified as follows :
\begin{equation}
\mathcal{C}\adj( \underline{\mathcal{A}} -\mathcal{A}_0  ) =0
\end{equation}
where $\mathcal{C}$ is a matrix with as many columns as constraints, and
 the columns of $\mathcal{C}$ are the same size as $\mathcal{A}$. The maximum
  of the EM functional with respect to $\underline \theta$ subject to the specified 
  linear constraints is then reached for:
\begin{equation}
\underline{\mathcal{A}} = \mathcal{M} - \mathcal{Q} \mathcal{C} \left( \mathcal{C}\adj \mathcal{Q} \mathcal{C} \right)\inv \mathcal{C}\adj( \mathcal{M} -\mathcal{A}_0  )
\end{equation}
and 
\begin{equation}
\underline R_{S,q} =  \textrm{diag} ( R_q^{ss})
\end{equation}
where ``$\textrm{diag}$'' returns a matrix with the same diagonal entries as its input argument.\\

\textbf{In the free noise case}, things are quite similar except
that the noise covariance matrices $R_{N,q}$ do not factorize out as nicely. 
The EM functional is again expressed as :
\begin{equation}
\Phi(\underline\theta , \theta) = -\frac{1}{2}  ( \underline{\mathcal{A}} - \mathcal{M} ) 
\mathcal{Q}(\underline{\mathcal{A}} - \mathcal{M} )\adj + const_{\underline A}
\end{equation}
where in this case:
\begin{equation}
\mathcal{Q} =  \sum_q n_q R_{N,q}\inv \otimes R_q^{ss}
\end{equation}
and
\begin{equation}
\mathcal{M} = \mathcal{Q} \inv \vect \left( \sum_q n_q R_{N,q}\inv R_q^{xs} \right)
\end{equation}
Then, the maximum of the EM functional with respect to $\underline \theta$ subject 
to the specified linear constraints is again reached for:
\begin{equation}
\underline{\mathcal{A}} = \mathcal{M} - \mathcal{Q} \mathcal{C} \left( \mathcal{C}\adj \mathcal{Q} \mathcal{C} \right)\inv \mathcal{C}\adj( \mathcal{M} -\mathcal{A}_0  )
\end{equation}
and 
\begin{equation}
\underline R_{S,q} =  \textrm{diag} ( R_q^{ss})
\end{equation}

These expressions of the re-estimates of the mixing matrix can become algorithmically 
very simple when for instance the linear constraints to be dealt with affect separate
 lines of $A$, or even simpler when the constraints are such that the entries of $A$ are affected separately.

\subsection*{Positivity constraints on the entries of $A$}

Suppose a subset of entries of $A$ are constrained to be positive. The maximization step of the 
EM algorithm on $A$ alone, again has to be modified. We suggest dealing with such constraints 
in a combinatorial way rephrasing the problem in terms of equality constraints. If the unconstrained 
 maximum of the EM functional is not in the specified domain, then one has to look for a maximum on 
 the borders of that domain: on a hyperplane, on the intersection of two, or three, or more hyperplanes.
One important point is that the maximum of the EM functional with respect to $A$ subject to 
a set of equality constraints will necessarily be lower than the maximum of the 
same functional considering any subset of these equality constraints. Hence, not all 
combinations need be explored, and a Branch and Bound type algorithm is well suited \cite{Nar}. A straightforward 
extension allows to deal with the case where a set of entries of the mixing matrix are constrained by upper and lower bounds.

\end{document}